\documentclass[reprint, amsmath, amssymb, aps, pra, showkeys]{revtex4-2}
\usepackage{xfrac}
\usepackage{amsmath}
\usepackage{bmpsize}
\usepackage{graphicx}
\usepackage{svg}
\usepackage{dcolumn}
\usepackage{bm}
\usepackage{blindtext}
\usepackage{braket}
\usepackage{times} 
\usepackage{lmodern} 
\usepackage{anyfontsize} 
\DeclareMathOperator{\sinc}{sinc}
\usepackage{bibunits}
\usepackage{nameref}
\usepackage[colorlinks=true, allcolors=blue]{hyperref}

\hypersetup{
    colorlinks=true,
    linkcolor=blue,
    filecolor=magenta,      
    urlcolor=cyan,
    citecolor=blue,
    pdftitle={Nasser et. al., On-Demand},
    pdfpagemode=FullScreen}
\usepackage[utf8]{inputenc} 
\DeclareUnicodeCharacter{A76C}{,} 
\usepackage[T1]{fontenc}
\usepackage{placeins}  
\FloatBarrier  

\begin{document}

\preprint{APS/123-QED}

\title{Erbium-Doped Fibre Quantum Memory for Chip-Integrated\\ Quantum-Dot Single Photons at 980~nm}

\author{Nasser Gohari Kamel$^{1,2}$}\thanks{These authors contributed equally.}
\author{Arsalan Mansourzadeh$^{1,2}$}\thanks{These authors contributed equally.}
\author{Ujjwal Gautam$^{1,2}$}
\author{Vinaya Kumar Kavatamane$^{1,2}$}
\author{Ashutosh Singh$^{1,2}$}
\author{Edith Yeung$^{3,4}$}
\author{David B. Northeast$^{4}$}
\author{Paul Barclay$^{1,2}$}
\author{Philip J. Poole$^{4}$}
\author{Dan Dalacu$^{3,4}$}
\author{Daniel Oblak$^{1,2}$}
\email{doblak@ucalgary.ca}

\affiliation{$^{1}$Institute for Quantum Science and Technology, University of Calgary, Calgary, AB, Canada T2N 1N4}
\affiliation{$^{2}$Department of Physics and Astronomy, University of Calgary, Calgary, AB, Canada T2N 1N4}
\homepage{https://qcloudlab.com/}
\affiliation{$^{3}$Department of Physics, University of Ottawa, Ottawa, Ontario, Canada K1N 6N5}
\affiliation{$^{4}$National Research Council Canada, Ottawa, ON, Canada K1A 0R6}

\date{\today}

\begin{abstract}

The realization of long-distance quantum communication and the envisioned quantum internet relies on coherent hybrid light–matter interfaces connecting quantum light emitters with quantum memory (QM) systems. Unlike probabilistic photon pair sources such as spontaneous parametric down-conversion, deterministic quantum light emitters enable the on-demand production of pure and bright single- and entangled- photons, essential for scalable quantum networks. In this work, we present the first experimental realization of a coherent hybrid light-matter interface between a chip-integrated InAsP/InP nanowire quantum dot (QD) and a solid-state QM based on $\text{Er}^{3+}$ ions doped in a glass silica fiber (erbium-doped fiber, EDF). The emission spectrum of the InAsP/InP nanowire QD aligns with the absorption bandwidth of the EDF at 980\,nm at cryogenic temperatures, allowing efficient interaction between the two systems. To demonstrate this, we present a spectroscopic characterization of the $^{4}I_{\sfrac{15}{2}} \leftrightarrow {}^{4}I_{\sfrac{11}{2}}$ optical transition in EDF at 980\,nm. Our measurements reveal substantial inhomogeneous broadening of this optical transition and a long spin population lifetime, underscoring EDF's potential for broadband QM implementation. We implement an 8\,GHz bandwidth multimode QM based on the Atomic Frequency Comb protocol, enabling the storage and retrieval of 59~weak coherent pulses. Furthermore, we characterize single-photon emission from an InAsP/InP nanowire QD at 980\,nm and demonstrate its deterministic storage and recall in the EDF QM. Notably, this is achieved without spectral tuning of the QD emission, demonstrating its direct compatibility with a solid-state QM.

\end{abstract}
\keywords{Quantum communication, Hybrid light-matter interface, Quantum memory, Quantum Dot}

\maketitle

\section{Main} \label{sec:Main}


The advancement of many photonic quantum technologies relies on the development of deterministic quantum light emitters and efficient light-matter interfaces~\cite{heindel2023quantum, o2009photonic, Uppu20211308}. These components are indispensable for long-distance quantum communication~\cite{Duan_2001, Lago_Rivera_2023} and quantum information processing~\cite{Ladd_2010, De_Raedt_2019}. While significant progress has been made in developing quantum light emitters across various physical platforms~\cite{aharonovich2016solid}, as well as photonic quantum memories (QMs)~\cite{heshami_2016, kamel-2025}, experimental realizations of hybrid light–matter interfaces between these systems remain relatively scarce~\cite{Thomas_2024, maruf_2023, Vural_2018, Chen_2016, Meyer_2015, Tang_2015, Akopian_2011}. A significant obstacle is attaining spectral compatibility between the QM's absorption wavelength and the quantum emitter's emission spectrum. Although the emission wavelength of quantum light emitters can be tuned through various techniques --- such as applying external electric or magnetic fields or inducing mechanical strain~\cite{Phoenix_2022, Lee_2017, maruf_2023} --- achieving long-term stability with these approaches remains challenging. Furthermore, the limited spectral tuning range presents a significant constraint on the practical realization of hybrid light–matter interfaces and hinders the implementation of scalable quantum networks~\cite{neuwirth2021quantum, ruskuc2024scalable}. \par

Among high-quality quantum light emitters~\cite{rodt_2020, aharonovich2016solid}, epitaxially grown InAsP/InP nanowire semiconductor quantum dots (InAsP QDs) stand out as a promising candidate due to their ability to generate bright and deterministic single and high-fidelity entangled photons~\cite{Versteegh_2014, Laferri_re_2022}. These QDs have also been integrated on silicon nitride waveguides, enabling hybrid single-photon sources on photonic chips~\cite{yeung2023chip}. Their emission wavelengths are determined primarily during the growth process, typically ranging from 880\,nm to 1000\,nm. Recent advancements have further extended InAsP QD emission into the telecom band~\cite{wakileh2024single, haffouz2020single}, enhancing their applicability for fiber-based quantum networks. Notably, these emission wavelengths are particularly advantageous for interfacing with various optical transitions of rare-earth ion-doped solids (REIDS)~\cite{carnall1989systematic} and the cesium D1-line~\cite{maruf_2023}, offering exciting prospects for hybrid light-matter interfaces. \par

Solid-state QMs based on REIDS offer a diverse range of optical transitions from visible to infrared and telecom-band~\cite{carnall1989systematic, tanabe1999optical}. This spectral versatility, combined with exceptionally long optical~\cite{Zhong_2015} and spin~\cite{Nicolas_2023} coherence times, makes REIDS a leading candidate for photonic QM applications. In addition, their wide inhomogeneous broadenings, ranging from tens of MHz to hundreds of GHz, enable robust, broadband, and highly multimode QM implementations~\cite{Wei_2024, Businger_2022, Ma_2021, Ortu_2022, Saglamyurek_2015}. Although REIDS-based quantum light storage has been successfully demonstrated using entangled photon pairs generated from probabilistic quantum light sources, such as spontaneous parametric down-conversion (SPDC) across a range of wavelengths~\cite{jiang2023quantum, Businger_2022, Lago-Rivera202137, puigibert2020entanglement}, the storage and retrieval of photons from deterministic quantum emitters remains significantly less explored. 

Achieving large brightness for deterministic quantum light emitters is associated with the short excited state lifetimes in solid-state emitters~\cite{aharonovich2014diamond} or fast radiative recombination of excitons (electron-hole pairs) in semiconductor QDs~\cite{Li2023, garcia2021semiconductor}. This unique characteristic of these emitters results in a large emission bandwidth (on the order of hundreds of MHz to GHz). This is also the case for SPDC-based sources, for which a high rate of entangled photon pair generation requires a high repetition rate of short pump laser pulses, resulting in high bandwidth photon emission from these sources~\cite{Couteau2018291}. 
Consequently, interfacing short temporal modes of the photons with QM requires the implementation of broadband and multimode QMs, with the Atomic Frequency Comb~\cite{Afzelius_2009} QM protocol being particularly well suited~\cite{Afzelius_2009, saglamyurek2011broadband}. 

To our knowledge, the only demonstration of solid-state QM for photons from a QD was reported by Tang \textit{et al.}~\cite{Tang_2015}. They achieved storage and retrieval of single photons collected directly from an InAs/GaAs QD via an objective lens, in an Nd$^{3+}$:YVO$_4$ crystal utilizing a 500\,MHz AFC QM. To accomplish this, the authors precisely engineered the QD's structural properties and identified—from hundreds of QDs—a single emitter whose wavelength matched the $\sim$\,500\,MHz inhomogeneously broadened $^{4}I_{9/2}\leftrightarrow{}^{4}F_{3/2}$ transition of Nd$^{3+}$:YVO$_4$. \par

In this study, we report the storage and retrieval of the negatively charged exciton emission, $\text{X}^{1-}$, from an InAsP nanowire QD integrated on a Si$_3$N$_4$ waveguide chip, in a QM based on the AFC protocol~\cite{Afzelius_2009}. The AFC is implemented on the 980\,nm optical transition of $^{4}I_{\sfrac{15}{2}}$ $\leftrightarrow$ $^{4}I_{\sfrac{11}{2}}$ in erbium-doped glass silica fiber. While the 980\,nm transition is commonly utilized as an optical pumping channel in erbium-based lasers and amplifiers~\cite{Brida2014409}, its coherence properties for quantum applications have not been explored to date. However, it should be noted that erbium-doped solids have been widely investigated for their telecom C-band transition, for which EDF exhibits significant inhomogeneous broadening and is a suitable platform for storing photonic qubits~\cite{Saglamyurek_2015, Wei_2024}. Before employing the 980\,nm transition of EDF for QM applications, we demonstrate its first spectroscopic investigation at cryogenic temperatures. We determine the spin population lifetime ($T_{1S}$) and optical coherence time ($T_{2O}$) of this transition. Due to the large inhomogeneous broadening of EDF, we achieve wavelength and bandwidth compatibility between the AFC at 980\,nm and the single photon emissions from the waveguide-coupled InAsP QDs, supporting effective hybrid light-matter interfacing. To prepare a multimode QM, we implement an 8\,GHz bandwidth AFC and demonstrate the storage and retrieval of 59 weak coherent temporal modes (laser probe pulses) as well as of the $\text{X}^{1-}$ emission from an InAsP QD.

 \begin{figure*}[htbp]
    \centering  \includegraphics[width=1\linewidth]{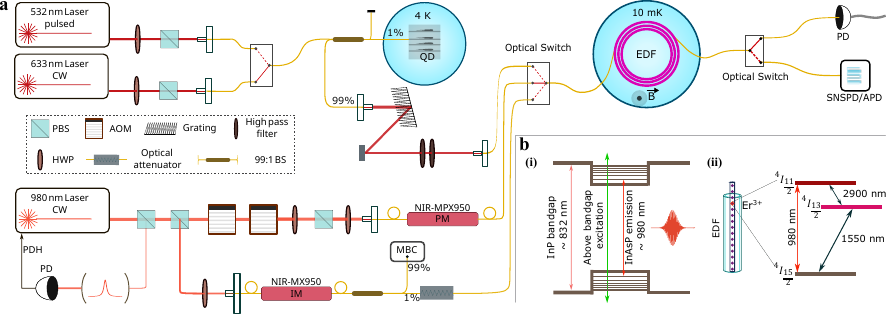}
    \caption{\textbf{Experimental setup, QD bandgap, and $\text{Er}^{3+}$ ion level structure.} \textbf{a,} Schematic of the experimental setup. A frequency-stabilized laser at 980\,nm is utilized for spin lifetime and optical coherence time measurements, as well as tailoring the inhomogeneous broadening for AFC combs and weak coherent pulse preparation during the storage and retrieval experiments. A CW He-Ne laser at 633\,nm, and a Toptica PicoTA picosecond laser at 532\,nm are utilized for continuous and pulsed, respectively, above-band excitation of the QD. All excitation sources are coupled to a 99:1 fiber beam splitter, where 1\% of the light is used for excitation. In the return path, we collect 99\% of all the light consisting of reflected excitation light, InP emission (Wurtzite emission at 832\,nm), and InAsP QD emission. See Supplementary Information for the electrical connections diagram of the experimental setup~\cite{Kamel_spp}. \textbf{b-i,} Bandgap structure of the InP nanowire and the tailored bandgap in the presence of the InAsP QD. \textbf{ii,} Energy level structure of the $\text{Er}^{3+}$ ion, illustrating transitions between the $^{4}I_{15/2}$, $^{4}I_{13/2}$, and $^{4}I_{11/2}$ states. Abbreviations: AOM, Acousto-Optic Modulators; APD, Avalanche Photodiode; BS, Beam Splitter; EDF, Erbium-doped glass silica fiber; HWP, Half-Wave Plate; IM, Intensity Modulator; PBS, Polarizing Beam Splitter; PM, Phase Modulator; PD, Photodetector; PDH, Pound–Drever–Hall laser locking system; SNSPD, Superconducting Nanowire Single-Photon Detector; QD, Quantum Dot.  } \label{fig:Er_Fiber_and_QD_Interface}
\end{figure*}

\begin{figure*}[htbp]
    \centering
    \includegraphics[width=0.93\linewidth]{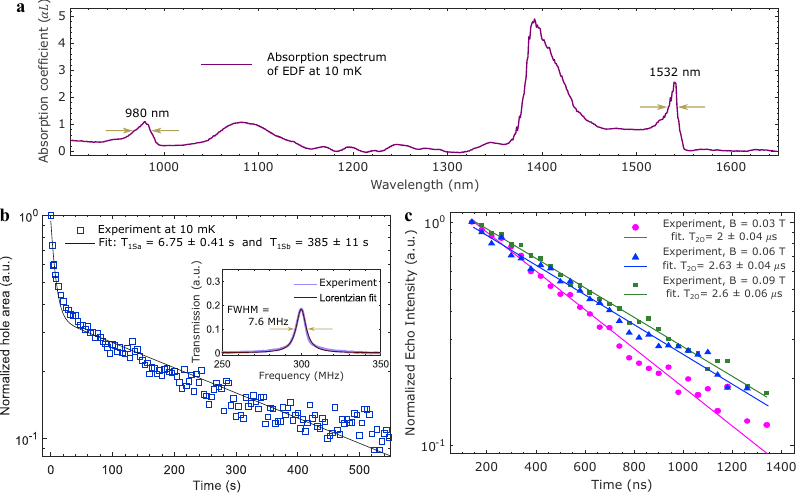}
\caption{\textbf{Spectroscopy of EDF for the 980\,nm transition.} \textbf{a,} Absorption spectra of EDF at 10\,mK and zero magnetic field where two Er transitions are marked at 980 and 1532\,nm. \textbf{b,} Spectral hole lifetime measured at 10\,mK with 0.09\,T magnetic field. The inset displays the first spectral hole recorded 50\,ms after the hole burning sequence. A Lorentzian fit reveals 7.58\,MHz linewidth of the narrowest spectral hole. Data is fitted with a double-exponential decay model, $A(t) = W_1 \text{exp}(-t/T_{1sa}) + W_2 \text{exp}(-t/T_{1sb})$, resulting spin lifetimes of $6.75\pm 0.414$\,s and $385\pm11$\,s with $W_1 = 0.59\pm0.04$ and $W_2 = 0.348\pm0.005$, respectively. \textbf{c,} Optical coherence times measured for three different magnetic fields with heterodyne detection. The results are fitted to a single exponential function as $I(t_{12})=I_{0}\text{exp}(-\frac{4t_{12}}{T_{2O}})$, where $I_{0}$ is the first echo intensity and $I(t_{12})$ is echo intensity at time separation $t_{12}$ between the $\pi/2$ and $\pi$ pulses. }\label{fig:Er_fiber_Absoption_hole_AFC_and_tempMode_storage}
\end{figure*}

\begin{figure*}[htbp]
    \centering
    \includegraphics[width=0.9\linewidth]{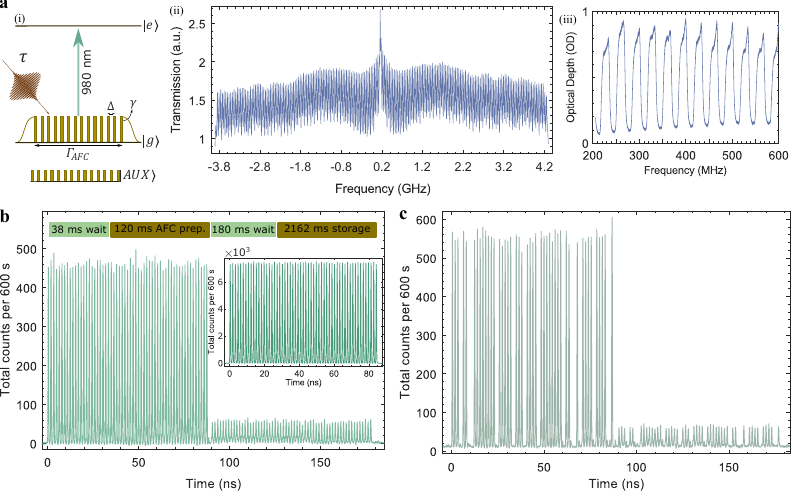}
    \caption{\textbf{AFC QM protocol implementation using EDF.} \textbf{a,} The process of generating periodic spectral holes is achieved through optical pumping, wherein electronic spins are selectively excited from the ground state, $\ket{g}$, to the excited state, $\ket{e}$. Subsequent spontaneous decays during long optical pumping transfers the spin population into an auxiliary state, $\ket{\text{AUX}}$, leading to the formation of persistent spectral absorbing features, as illustrated in the schematic (i). The prepared AFC comb can store multimode photonic qubits with a duration of $\tau_{\Gamma}=2.5/\Gamma_{\text{AFC}}$. (ii) An 8\,GHz bandwidth AFC is prepared with a storage time of 30\,ns, demonstrating squarish absorbing teeth. (iii) A zoomed view of the AFC provides a detailed 400\,MHz scan for enhanced visualization. \textbf{b,} Storage and recall of 59 weak coherent temporal modes. The inset represents the input probe pulses before interacting with the sample. The mean photon number per mode is $\mu = 1.15 \times 10^{-4}$. Here, the temporal duration of each mode is 320\,ps, and 1.12\,ns is the time delay between them. The colored band demonstrates the experimental timeline for different stages. The storage sequence begins with a 38\,ms wait to ensure optical switches are set for AFC preparation. AFC preparation pulses are applied for 120\,ms, followed by a 180\,ms wait to allow spontaneous decay of the excited state and to reconfigure the optical switches for single-photon storage. Each sequence lasts 2.5\,s and is repeated 240 times to accumulate the storage histogram. \textbf{c,} Here, weak coherent probe pulses are randomly distributed over 85\,ns and are stored in the AFC with 90\,ns storage time. The temporal profile of recalled modes is perfectly preserved, matching that of the input modes. See Supplementary Information for more details of the AFC preparation method and storage~\cite{Kamel_spp}. } 
\label{fig:Random_Temporal_modes}
\end{figure*}

\begin{figure*}[htbp]
\centering
\includegraphics[width=1\linewidth]{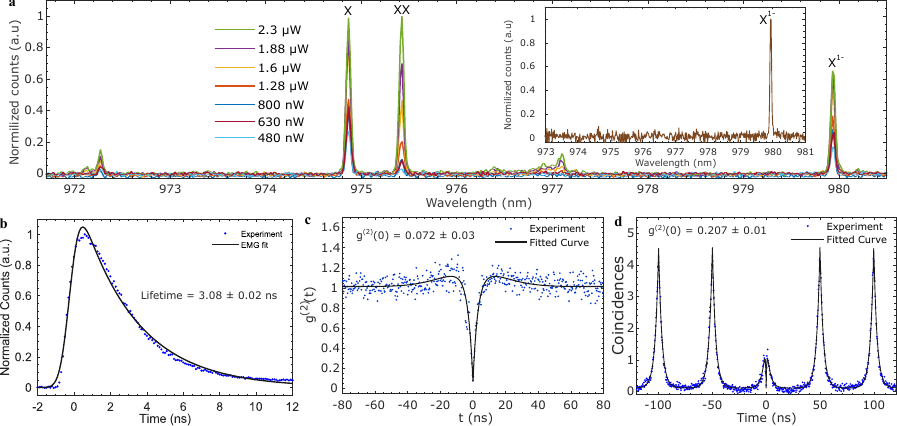}
\caption{\textbf{InAsP quantum dot emission characteristics.} \textbf{a,} Full emission spectrum of the QD under varying excitation powers using a CW 633\,nm laser for above-band excitation. The inset represents the negatively charged exciton, $\text{X}^{1-}$, filtered with a 1200 grooves/mm grating and two free space high pass filters. \textbf{b,} A lifetime of $\tau = 3.08$\,ns corresponding to $\text{X}^{1-}$ emission is measured using above-band excitation by a 532\,nm picosecond pulsed laser operating at 10\,MHz repetition rate. The lifetime is extracted by fitting the results to an exponentially modified Gaussian (EMG) function described in~\cite{maruf_2023}. Second-order autocorrelation measurements of the filtered $\text{X}^{1-}$ emission $g^{(2)}(t)$, under CW, \textbf{c}, and pulsed excitations, \textbf{d}, yielding $g^{(2)}(0) = 0.072$ and coincidences in the central peak around $\tau=0$ measure 0.207, respectively. The $g^{(2)}$ results for CW and pulsed excitation are measured over 7.5 hours and are fitted according to models described in~\cite{stachurski2022single, holewa2024high}. With the pulsed $g^{(2)}$ measurements, we have subtracted the background corresponding to detector dark counts and random excitations due to pulsed laser leak light.}
\label{fig:QD_emission_filtered_lifetime_G2}
\end{figure*}

\section{Results}\label{sec:Results}

The experimental setup, illustrated in FIG.~\ref{fig:Er_Fiber_and_QD_Interface}\textbf{\textcolor{blue}{a}}, consists of a 10\,m-long glass silica fiber doped with 200~ppm natural abundance $\text{Er}^{3+}$ ions, cooled to $\sim$\,10\,mK. The EDF is placed inside a superconducting solenoid, allowing the application of tunable magnetic fields to induce Zeeman splitting of the $\text{Er}^{3+}$ ion energy levels. The semiconductor InAsP QDs embedded in InP nanowires are located on the 4\,K stage in the same cryostat. The nanowires are interfaced with Si$_3$N$_4$ photonic waveguides, allowing evanescent coupling of light between the nanowires and waveguides~\cite{Yeung_2023} (see 4\,K flange schematic in FIG.~\ref{fig:Er_Fiber_and_QD_Interface}\textbf{\textcolor{blue}{a}}). 
FIG.~\ref{fig:Er_Fiber_and_QD_Interface}\textbf{\textcolor{blue}{b-i}} illustrates the band structure of the InAsP QD embedded within an InP nanowire, while FIG.~\ref{fig:Er_Fiber_and_QD_Interface}\textbf{\textcolor{blue}{b-ii}} presents the energy level diagram of the $\text{Er}^{3+}$ ions. This highlights that the QD emission is spectrally aligned with the $^{4}I_{\sfrac{15}{2}} \leftrightarrow {}^{4}I_{\sfrac{11}{2}}$ transition of $\text{Er}^{3+}$, providing a solid foundation for coupling semiconductor QDs with solid-state QMs. Additional details on the experimental setup and the subsequent experiments are provided in the Methods section~\ref{sec:Methods}. \par

The spectroscopic characterizations of the 980\,nm transition of EDF are presented in FIG.~\ref{fig:Er_fiber_Absoption_hole_AFC_and_tempMode_storage}. The EDF absorption spectrum is measured at $\sim$\,10\,mK and zero magnetic field, for which a 10\,nm (FWHM) inhomogeneous broadening of the 980\,nm transition is observed (see FIG.~\ref{fig:Er_fiber_Absoption_hole_AFC_and_tempMode_storage}\textbf{\textcolor{blue}a}). For wavelengths $\lambda \leq 1150$\,nm, the single-mode fiber (SMF28) used to connect to the EDF in- and output- can guide a second transverse mode, while the smaller core size of the EDF still supports only a single mode. This induces additional loss at the interface of the SMF28 and EDF fiber, creating the appearance of background absorption below $\sim 1150$\,nm, compared to the off-resonant range at $\lambda \geq 1540$\,nm, where the absorption coefficient is effectively zero. Additionally, the absorption spectrum features distinct peaks at 1080\,nm and 1400\,nm, which are characteristic of silica optical fibers.\par


To investigate the time-resolved population dynamics, $T_{1S}$, of the Zeeman-split spin-levels in $\text{Er}^{3+}$, utilizing a narrow linewidth laser a persistent spectral hole ($\text{FWHM} = 7.58$\,MHz) is created at a magnetic field of 0.09\,T (see Methods for more details). The optical pumping achieves 86\% spin population transfer efficiency relative to the peak optical depth (OD), resulting in the formation of persistent deep spectral holes. The optical depth of the spin-population remaining is denoted as $d_0$.
The hole dynamics are monitored 50\,ms after optical pumping by measuring the persistent hole area as a function of time. The results, depicted in FIG.~\ref{fig:Er_fiber_Absoption_hole_AFC_and_tempMode_storage}\textbf{\textcolor{blue}{b}}, exhibit a biexponential decay of $T_{1Sa}=6.8$\,s and $T_{1Sb}=6.4$\,min, indicating the presence of two distinct ion classes. A similar behavior has been observed and extensively analyzed for the 1532\,nm transition in the same EDF sample~\cite{bornadel2024hole, Saglamyurek_2015_hole}. The measured long $T_{1Sb}$ verifies the possibility of spectrally tailoring the absorption profile as required for multiple photon-echo protocols. \par

In addition, employing the standard Hahn echo~\cite{hahn1950spin} method, the optical coherence time, T$_{2O}$, of the 980\,nm transition in EDF is measured at three magnetic fields where we observe a coherence time of $\sim$\,2.6$\,\mu$s at $B=0.06 \text{ and }0.09$\,T, and $\sim$\,2$\,\mu$s at $B=0.03$\,T (see FIG.~\ref{fig:Er_fiber_Absoption_hole_AFC_and_tempMode_storage}\textbf{\textcolor{blue}{c}}). This optical coherence time, in conjunction with large inhomogeneous broadening and long spin population lifetime, affirms that the 980\,nm transition in the EDF is suitable for implementing AFC QM, suitable for storing broadband photonic modes. \par

Here, we realize an AFC on the 10\,nm inhomogeneously broadened 980\,nm transition of the EDF. With the AFC QM protocol (schematically presented in the FIG.~\ref{fig:Random_Temporal_modes}\textbf{\textcolor{blue}{a-i}}), the inhomogeneous broadening of the atomic ensemble is spectrally tailored by optical pumping to form periodic absorption comb-like features with a tooth spacing $\Delta$ in the frequency domain. The resulting AFC enables storage of photons for a time given by $t_\mathrm{s}=1/\Delta$. For an AFC QM the total bandwidth, $\Gamma_{\mathrm{AFC}}$, is defined as $\Gamma_{\mathrm{AFC}} = N \times \Delta$, where $N$ is the number of absorbing teeth, and $F$ is the finesse of the AFC. The finesse, in turn, is defined as $F = \frac{\Delta}{\gamma}$, where $\gamma$ represents the linewidth of each absorbing tooth.  

To prepare the AFC combs, we introduce an approach involving the modulation of both the amplitude and phase of the optical-pump pulses, which effectively tailors the inhomogeneous broadening into squarish-shaped AFC comb teeth. Compared to conventional techniques, as described in~\cite{Businger_2022}, our approach demonstrates approximately an order of magnitude improvement in spectral hole burning efficiency (see Supplementary Information for additional details~\cite{Kamel_spp}). For all storage experiments, we create 8\,GHz wide AFC combs with different predetermined storage times from 5 to 100\,ns. Figure~\ref{fig:Random_Temporal_modes}\textbf{\textcolor{blue}{a-ii}} illustrates an example of an 8\,GHz AFC prepared for 30\,ns storage time. For improved clarity, the optical depth corresponding to a 400\,MHz zoomed-in segment of the AFC is shown in Figure~\ref{fig:Random_Temporal_modes}\textbf{\textcolor{blue}{a-iii}}.   \par

\begin{figure*}[htbp]
    \centering
\includegraphics[width=0.9\linewidth]{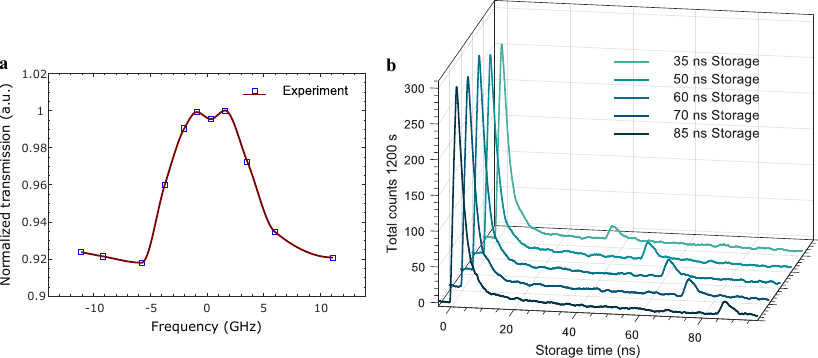}
\caption{\textbf{Storage of $\text{X}^{1-}$ emission in 980\,nm transition of $\text{Er}^{3+}$ ion.} \textbf{a,} To achieve spectral overlap between the $\text{X}^{1-}$ emission and the AFC bandwidth in the 980\,nm transition of EDF, we create a 8\,GHz broad spectral hole and transmit $\text{X}^{1-}$ single photons through the EDF while monitoring changes in transmitted intensity. Since the frequency of QD emission remains fixed throughout the experiment, we systematically vary the center frequency of the laser to burn spectral holes at different regions. An increase in transmission indicates successful bandwidth matching between the $\text{X}^{1-}$ emission and the final AFC combs. \textbf{b,} Storage of $\text{X}^{1-}$ in EDF for five different storage times.}\label{fig:QD_laser_bandwidth_match_and_QD_storage}
\end{figure*}  

The efficiency of the AFC QM protocol is described by 
\begin{equation}
    \eta_{\text{AFC}} = \frac{d^2}{F^2} \exp(-\frac{d}{F})\sinc^2 \left( \frac{\pi}{F} \right)e^{-d_0},
    \label{eq:AFC_eff}
\end{equation}
where $d$ is the OD before spectral tailoring~\cite{li-2025}. The efficiency degrades with background OD, $d_0$, as $\eta_{\text{AFC}} \sim \text{e}^{-d_0}$. It has been shown that the most efficient AFC QMs have squarish-absorbing features~\cite{li-2025}, while the optical pumping sequence effectively transfers the spin populations of the periodic spectral holes, resulting in the background OD being close to zero. 
Our EDF sample exhibits a maximum OD of 1.1 on the 980\,nm transition. We tailor AFCs with squarish absorbing teeth, a background OD of 0.05, and a finesse of $F \simeq 2$, resulting in an expected memory efficiency of 6.7\%. However, due to 83\% transmission loss, primarily caused by splicing at both ends of the EDF to single-mode fibers and connector losses, the measured total memory efficiency for storing and recalling weak coherent pulses is 1\%, as expected.\par

The multimode capacity of our QM is shown in FIG.~\ref{fig:Random_Temporal_modes}\textbf{\textcolor{blue}{b}}, as we simultaneously store and recall 59~weak coherent temporal modes for 90\,ns. The temporal duration of each mode, $\tau =320$\,ps, is close to the smallest value of $\tau_\Gamma = {2.5}/{\Gamma_{AFC}}$ compatible with the AFC bandwidth~\cite{Businger_2022}. Therefore, for $\Gamma_{AFC}=8$\,GHz and a storage time of 90\,ns, it is possible to efficiently store 144 temporal modes, assuming 312.5\,ps for both temporal duration and a spacing between the modes.
However, due to limitations of the bandwidth of our detection system, we choose 1.12\,ns temporal mode separation, resulting in a lower number of modes stored.\par 

In extensive experimental implementations of AFC-based QMs, this protocol has been shown to re-emit stored photonic qubits with fidelities close to unity~\cite{saglamyurek2011broadband, Businger_2022}. To evaluate the fidelity of our QM for time-bin encoding, FIG.~\ref{fig:Random_Temporal_modes}\textbf{\textcolor{blue}{c}} presents the storage and recall of randomly encoded weak coherent temporal modes~\cite{usmani-2010}. The retrieved modes are re-emitted in the same order as they were stored, with no observable cross-talk between the temporal modes. Hence, the fidelity of the time-bin storage is purely limited by background noise. Given the signal-to-noise ratio (SNR) of 6.08 for the recalled weak coherent pulses, we assess that the storage fidelity of a time-bin qubit would be $F_{q} = 87.6$\%. This clearly exceeds the maximal fidelity bound achievable with a classical storage strategy of $F_{c}^{(max)}=0.667$, which takes into account the mean photon number of $\mu=1.15\times10^{-4}$ at the input and total memory efficiency of 1\% \cite{gundougan2012quantum}, confirming the quantum nature of the storage process. The SNR is primarily limited by the 200\,Hz dark counts of the SNSPD.
 \par

Next, we present the characteristics of the InAsP QD used in this study. FIG.~\ref{fig:QD_emission_filtered_lifetime_G2}\textbf{\textcolor{blue}{a}} shows the QD emission spectra with above-band excitation using a 633\,nm CW He-Ne laser at varying excitation powers. The three main emissions of neutral exciton, $\text{X}$, biexciton cascade, $\text{XX}$, and negatively charged exciton, $\text{X}^{1-}$, are clearly observed in this plot with their emission rate increasing with excitation power. At excitation powers exceeding 1.28\ensuremath{\,}$\mu$W, additional emission features appear around 972\,nm and 977\,nm, attributable to different charge carrier recombination processes within the QD spectrum. Despite these unwanted peaks, due to limited collection efficiency and filtering losses, all subsequent experiments were conducted with an excitation power of more than 1.7\ensuremath{\,}$\mu$W (above saturation power).  \par

The $\text{X}^{1-}$ emission of this particular InAsP QD is spectrally aligned with the center of the absorption spectrum of the EDF. To enable further characterization, we selectively filter the $\text{X}^{1-}$ emission, as it is well spectrally separated from both the $\text{X}$ and $\text{XX}$ emissions (see inset of the Figures~\ref{fig:QD_emission_filtered_lifetime_G2}\textbf{\textcolor{blue}{a}}). Figure~\ref{fig:QD_emission_filtered_lifetime_G2}\textbf{\textcolor{blue}{b}} presents the corresponding measured lifetime of the $\text{X}^{1-}$ emission, and \ref{fig:QD_emission_filtered_lifetime_G2}\textbf{\textcolor{blue}{c}}, and \ref{fig:QD_emission_filtered_lifetime_G2}\textbf{\textcolor{blue}{d}} are the second-order autocorrelation function measurement ($g^{(2)}(t)$)~\cite{Laferriere2021} for CW and pulsed excitation, respectively. In our experiments, above-band excitation and the effect of excitation power exceeding saturation power results in the creation of additional charge carriers (re-excitation) such that we observe a slightly longer lifetime compared to the expected value of 1-2\,ns, as reported in~\cite{Yeung_2023,maruf_2023}. This also has a clear effect on the $g^{(2)}(t)$ and coincidence measurements with rising and falling peaks around zero delay~\cite{laferriere2020multiplexed}. In addition, due to the short lifetimes of InAsP QDs, their emission is typically lifetime-limited to spectral bandwidths from hundreds of MHz to a few\,GHz~\cite{Yeung_2023, maruf_2023}. \par

Finally, we demonstrate the implementation of a hybrid light-matter interface by storing and retrieving $\text{X}^{1-}$ emission matching the 980\,nm transition of the EDF. The first task is to precisely overlap the wavelength of the $\text{X}^{1-}$ emission to that of the AFC. To that end, we first employ a spectrometer with 50\,GHz resolution to measure and coarsely align the wavelength of the 980\,nm laser used for AFC preparation to the $\text{X}^{1-}$ emission wavelength. Next, we use the laser to create an 8\,GHz wide spectral hole in the EDF and monitor the transmission of the $\text{X}^{1-}$ emission through the fiber with an Avalanche Photodiode (APD), while we fine-tune the wavelength of the laser. When the wavelength of the QD overlaps with the spectral hole created by the optical pump laser, an increase in counts is observed on the APD as shown in FIG.~\ref{fig:QD_laser_bandwidth_match_and_QD_storage}\textbf{\textcolor{blue}{a}}. Although the transmitted intensity of the $\text{X}^{1-}$ can be fitted with a Lorentzian function yielding a FWHM of 8.05\,GHz, this value does not correspond to the intrinsic linewidth of this emission. Instead, the emission linewidths of quantum light emitters can be measured with either a high-resolution single-photon spectrometer or with scanning Fabry-Perot cavities. \par

FIG.~\ref{fig:QD_laser_bandwidth_match_and_QD_storage}\textbf{\textcolor{blue}{b}} presents the storage and retrieval of $\text{X}^{1-}$ emission from the InAsP QD in the EDF QM for five different storage durations under a 0.06\,T magnetic field. The QD is optically excited using a 532\,nm picosecond laser operating at a 10\,MHz repetition rate, while the AFC preparation, storage sequence, and timing remain identical to those used for weak coherent pulse storage (as detailed in FIG.~\ref{fig:Random_Temporal_modes}\textbf{\textcolor{blue}{b}}). Due to the low photon extraction efficiency from the QD, the results are recorded over a 1200\,s integration time (480 repetitions of the storage sequence). Here, with 1\% QM efficiency, we observe that the temporal profile of the recalled photons is well preserved. We measure an SNR of 1.92 for the retrieved $\text{X}^{1-}$ emission, which is mainly limited by the 200 dark counts per second from the SNSPD. \par
 
In our study, we do not encode a qubit into the $\text{X}^{1-}$ single photons; instead, we directly map the temporal profile of the single photons onto the QM. We measure the pulsed second-order correlation function of $g^{(2)}_{\text{in}}(0) = 0.207\pm0.01$ for the $\text{X}^{1-}$ emission prior to storage over 7.5 hours. Due to limited collection efficiency from the chip-integrated QD, 83\% transmission loss, and QM efficiency, we present here an estimate of the $g^{(2)}_{\text{out}}(0)$ for recalled photons, using the method described in~\cite{Thomas_2024}, given by 
\begin{equation}
    g^{(2)}_{\text{out}}(0) = \frac{1 + S^2 + g^{(2)}_{\text{in}}(0)}{(1+ S)^2},
\end{equation}
where $S$ denotes the SNR. Given an SNR of 1.92, we obtain $g^{(2)}_{\text{out}}(0) = 0.5547 \pm 0.001$, which remains below the classical limit of 1.\par

\section{Discussion and Outlook}\label{subsubsec:Discussion_and_outlook}

In this research, we have displayed the potential of the 980\,nm optical transition of $\text{Er}^{3+}$ ions doped in glass silica fibers for advanced quantum storage applications. We have characterized this transition, demonstrating spin population dynamics toward the realization of long-lived AFC combs for photonic qubit storage and the storage of 59 weak coherent temporal modes. This marks a significant step toward discovering robust QMs compatible with hybrid quantum architectures for long-distance quantum communication and quantum processing applications.\par

We characterized the emission of InAsP QDs and utilized $\text{X}^{1-}$ emitted photons to demonstrate their successful storage and recall within 8\,GHz-wide AFCs prepared in EDFs for five distinct storage times. To ensure precise bandwidth matching between the $\text{X}^{1-}$ emission and the prepared AFC, we employed a combination of spectrometer for rough tuning between $\text{X}^{1-}$ emission and 980\,nm laser and hole burning technique to verify precise bandwidth matching of the memory and QD emission. The large inhomogeneous broadening inherent to EDFs has been harnessed as a critical resource for bridging the interface between InAsP QD single photon source and a QM. This achievement exemplifies the viability of EDFs as QM platforms for interfacing with on-demand quantum light sources.\par

Looking forward, the seamless integration of EDFs into existing fiber-based quantum communication infrastructures highlights their practicality. However, further progress in this platform depends on the fabrication of isotopically purified rare-earth doped fibers to minimize decoherence effects. Additionally, removing host materials that contain nuclear spins could significantly improve coherence properties and enhance the storage efficiency of rare-earth ion-doped fibers. These advancements would enable more scalable and high-performance quantum communication networks.  \par

Beyond EDFs, continued progress in semiconductor QD growth, particularly towards highly tunable emission, will be crucial for realizing more sophisticated hybrid light-matter interactions. In this context, $^{171}\text{Yb}^{3+}:\text{Y}_2\text{SiO}_5$, with its optical transition at 978.54\,nm and remarkable optical and spin coherence properties~\cite{Nicolas_2023}, represents another promising candidate for hybrid interfacing with InAsP QDs. Furthermore, exploring other optical transitions in rare-earth ions through similar spectroscopic studies could open new possibilities for developing advanced hybrid light-matter interfaces.

\section{Methods}  \label{sec:Methods}

In this study, we employ a 10\,m-long EDF with an erbium ion concentration of 200 ppm. The fiber is cooled to the cryogenic temperature of 10\,mK using a BlueFors dilution refrigerator and positioned within a superconducting magnet capable of generating magnetic fields up to 2 T. To ensure uniform thermal distribution, the fiber was spooled around a 4\,cm diameter oxygen-free copper cylinder, with cryogenic gel applied between layers to enhance thermal conductivity and maintain consistent temperature throughout the fiber. The absorption spectra of EDF at 10\,mK is measured using a broadband light source and an Agilent optical spectrum analyzer. Here, the intensity vs wavelength is recorded before and after tEDF as $I_{\text{in}}$ and $I_{\text{out}}$ which according to the Beer–Lambert law, $I_{\text{out}} = C I_{\text{in}}\text{exp}(-\alpha L)$, the absorption coefficient, $\alpha L$ is extracted. The parameter $C$ accounts for all the losses, including splicing and connectors.    \par

In the experimental setup (see FIG.~\ref{fig:Er_Fiber_and_QD_Interface}), the CW output, at centre frequency \( \nu_0 \), of a frequency stabilized, Toptica DLC pro, 980\,nm laser (linewidth less than 70\,kHz) is split into two arms. In one arm, we use two acousto-optic modulators (AOMs, AA Opto-Electronic: MT200-B100A0,5-1064 and Brimrose: TEM-85-10) in series to gate the CW light while realizing the amplitude modulation required to create pulses for spectroscopy and AFC tailoring. Additional neutral density filters are utilized to suppress any remaining leak light. After the two AOMs, the laser frequency will shift to \( \nu_0 +285\)\,MHz. To shift or sweep the laser frequency, an electro-optic phase modulator (PM, IxBlue: NIR-MPX950) is employed. This beam path is used for spectral hole burning, AFC comb preparation, and optical coherence time measurements. \par         

The second arm is derived from the 0'th order beam from the AOM in the first arm, with about 2\,mW coupled into an optical fiber. It is utilized to prepare weak coherent pulses with a duration of 320 ps at the center frequency \( \nu_0 \) using an electro-optic intensity modulator (IM, IxBlue: NIR-MX950) and subsequently attenuated to achieve a mean photon number of \( \mu = 1.1515 \times 10^{-4} \), corresponding to a single photon detection probability of \( P(1) = 1.151 \times 10^{-4} \) per pulse. The mean photon number is attained from the number of photons detected over the number of applied weak coherent pulses before the cryostat. Since the total shift frequency after two AOMs is 285\,MHz, we prepare the 8\,GHz AFC combs such that this frequency offset between weak coherent pulses and the center of the AFC combs is compensated. Therefore, the inhomogeneous broadening tailored from \( \nu_0 -3.8\)\,GHz to \( \nu_0 +4.2\)\,GHz overlaps the AFC bandwidth with the weak coherent pulses. A 99:1 beam splitter is used after the IM, directing 99\% of the light to a modulation bias controller (MBC) for precise laser intensity control. The IM's DC bias is locked and adjusted in real time to minimize transmission. The remaining 1\% of the light is further attenuated to effectively suppress any residual leakage. Due to the IM's limited extinction ratio of 25\,dB, an additional attenuation of 100\,dB is applied to fully eliminate leaked light from the 1\% output of the beam splitter, preventing any unintended interference between the recalled echoes and leak lights.\par

We investigate the spin population dynamics, $T_{1S}$, of EDF at 980\,nm by creating the narrowest possible spectral hole and analyzing its time-resolved hole area. To achieve spectral hole burning, optical pulses are generated through AOMs at the center frequency of \( \nu_0 +285\). Furthermore, this frequency is shifted with PM by 300\,MHz, where we employ a serodyne RF signal resulting in a single sideband shift of the laser frequency from to \( \nu_0 + 585 \)\,MHz. This approach selectively enhances the desired sideband while minimizing the power in the zeroth-order and first negative order components. The main reason for shifting the laser frequency is to prevent any interaction between the spins in the spectral hole region and residual light at a frequency of \( \nu_0+285 \). Given the long spin lifetime, we create only a single spectral hole throughout the experiment and monitor its population dynamics. Any residual light at \( \nu_0 + 285 \)\,MHz can lead to depopulation of the hole, thereby compromising the accuracy of our measurements of the spin population lifetime. After burning the spectral hole, we begin monitoring its dynamics using a weak probe signal scanned over a broad range (\( \nu_0 + 400 \)\,MHz to \( \nu_0 + 800 \)\,MHz). The decay dynamics reveal two distinct exponential components, indicating the existence of two different ion classes~\cite{bornadel2024hole}. 

To measure the optical coherence time, $T_{2\text{O}}$, two short optical pulses with a duration of 25\,ns and 50\,ns with a variable time separation between them and resonant with $^{4}I_{\sfrac{15}{2}} \leftrightarrow {}^{4}I_{\sfrac{11}{2}}$ transition are employed as $\pi/2$ and $\pi$ pulses, respectively. The resulting echo amplitude is detected using a heterodyne detection method, employing a local oscillator with a center frequency shifted by 85\,MHz relative to the $\pi/2$ and $\pi$ pulses. Fast Fourier Transform (FFT) analysis of the heterodyne beat signal is performed to extract the $T_{2\text{O}}$. Each data point in FIG.~\ref{fig:Er_fiber_Absoption_hole_AFC_and_tempMode_storage}\textbf{\textcolor{blue}{c}} represents an average of twenty time measurements for a particular time delay between the $\pi/2$ and $\pi$ pulses. \par

Inside the same cryostat, the chip-integrated InAsP QDs~\cite{Yeung_2023} are mounted on the 4\,K flange. The QDs are optically excited using either CW or picosecond pulsed lasers for above-band excitation. Excitation and emission collection are achieved through a single-mode optical fiber in an edge-coupling configuration. This fiber, mounted on a nanopositioner, enables precise alignment with the target Si$_3$N$_4$ waveguide on the photonic chip, enabling evanescent coupling of the excitation light and emitted photons between QD and waveguide. Although QD emission ideally transmits symmetrically from both sides of the nanowire (approximately 50\% in each direction), we collect the emissions only from one side, resulting in an estimated post-filtering collection efficiency of about 0.1\%. The primary sources of loss include evanescent coupling to the Si$_3$N$_4$ waveguide, fiber-waveguide coupling, and free-space optical elements used for filtering.  \par

Outside the cryostat, a 99:1 beam splitter directs the majority of the QD emission toward two optical filtering setups. The first setup (not shown in the experimental schematic FIG~\ref{fig:Er_Fiber_and_QD_Interface}a) used to capture the entire emission spectrum in FIG.~\ref{fig:QD_emission_filtered_lifetime_G2}a) employs only two free-space high-pass filters in series to isolate the full QD emission bandwidth while suppressing unwanted light, such as Wurtzite-related emission from the InP bandgap at 832\,nm and residual excitation laser reflections. The second setup (shown in FIG~\ref{fig:Er_Fiber_and_QD_Interface}a), used for all correlation and storage experiments, consists of a high-resolution filtering system that also incorporates a 1200 grooves/mm grating with a spectral resolution of 0.5\,nm in addition to the two free-space high-pass filters. This selectively isolates the 980\,nm emission of the negatively charged exciton state ($\text{X}^{1-}$) from all other emission lines of the QDs.\par

The storage protocol, as illustrated in FIG.~\ref{fig:Random_Temporal_modes}\textbf{\textcolor{blue}{b}}, involves a 38\,ms wait to ensure the optical Micro-Electro-Mechanical Systems (MEMS) switch is activated for AFC burn pulses. The AFC comb preparation (see Supplementary Information for amplitude and phase modulated optical pulses~\cite{Kamel_spp}) is then performed for 120\,ms, followed by a 180\,ms wait to ensure complete decay of the excited state population, and the MEMS switch to activate for transmission of weak coherent pulses or $\text{X}^{1-}$. Finally, the prepared photonic modes (weak coherent pulses or $\text{X}^{1-}$ emission) are stored and recalled with a 4\,MHz rate for 2162\,ms in each cycle, resulting in a 2500\,ms total duration of one storage sequence. This sequence was repeated 240 (480) times while SNSPD detection histograms were captured for 600\,s (1200\,s) during storage of weak coherent or $\text{X}^{1-}$ emission, respectively. At the output of the EDF, we use another optical MEMS switch to prevent disabling the SNSPDs with high-power AFC comb preparation laser pulses. In addition, by electrical gating, we turn off the SNSPD bias current for the first 338\,ms of the storage sequence.

\begin{acknowledgments}
The authors would like to thank Amir Ahadi, Leili Esmaeilifar, and Anuj Sethia for assistance with the experimental setup. This work was supported by the National Research Council of Canada through the Small Teams Initiative QPIC, the Government of Alberta Major Innovation Fund Project on Quantum Technologies, Alberta Innovates Advance Program, the Canadian Foundation for Innovation Infrastructure Fund (CFI-IF), and the Natural Sciences and Engineering Research Council of Canada (NSERC) through the Alliance Quantum Consortia Grants CanQuest and ARAQNE. 
\end{acknowledgments}

\bibliography{MainReferences1}

\begin{thebibliography}{58}%
\makeatletter
\providecommand \@ifxundefined [1]{%
 \@ifx{#1\undefined}
}%
\providecommand \@ifnum [1]{%
 \ifnum #1\expandafter \@firstoftwo
 \else \expandafter \@secondoftwo
 \fi
}%
\providecommand \@ifx [1]{%
 \ifx #1\expandafter \@firstoftwo
 \else \expandafter \@secondoftwo
 \fi
}%
\providecommand \natexlab [1]{#1}%
\providecommand \enquote  [1]{``#1''}%
\providecommand \bibnamefont  [1]{#1}%
\providecommand \bibfnamefont [1]{#1}%
\providecommand \citenamefont [1]{#1}%
\providecommand \href@noop [0]{\@secondoftwo}%
\providecommand \href [0]{\begingroup \@sanitize@url \@href}%
\providecommand \@href[1]{\@@startlink{#1}\@@href}%
\providecommand \@@href[1]{\endgroup#1\@@endlink}%
\providecommand \@sanitize@url [0]{\catcode `\\12\catcode `\$12\catcode
  `\&12\catcode `\#12\catcode `\^12\catcode `\_12\catcode `\%12\relax}%
\providecommand \@@startlink[1]{}%
\providecommand \@@endlink[0]{}%
\providecommand \url  [0]{\begingroup\@sanitize@url \@url }%
\providecommand \@url [1]{\endgroup\@href {#1}{\urlprefix }}%
\providecommand \urlprefix  [0]{URL }%
\providecommand \Eprint [0]{\href }%
\providecommand \doibase [0]{https://doi.org/}%
\providecommand \selectlanguage [0]{\@gobble}%
\providecommand \bibinfo  [0]{\@secondoftwo}%
\providecommand \bibfield  [0]{\@secondoftwo}%
\providecommand \translation [1]{[#1]}%
\providecommand \BibitemOpen [0]{}%
\providecommand \bibitemStop [0]{}%
\providecommand \bibitemNoStop [0]{.\EOS\space}%
\providecommand \EOS [0]{\spacefactor3000\relax}%
\providecommand \BibitemShut  [1]{\csname bibitem#1\endcsname}%
\let\auto@bib@innerbib\@empty
\bibitem [{\citenamefont {Heindel}\ \emph {et~al.}(2023)\citenamefont
  {Heindel}, \citenamefont {Kim}, \citenamefont {Gregersen}, \citenamefont
  {Rastelli},\ and\ \citenamefont {Reitzenstein}}]{heindel2023quantum}%
  \BibitemOpen
  \bibfield  {author} {\bibinfo {author} {\bibfnamefont {T.}~\bibnamefont
  {Heindel}}, \bibinfo {author} {\bibfnamefont {J.-H.}\ \bibnamefont {Kim}},
  \bibinfo {author} {\bibfnamefont {N.}~\bibnamefont {Gregersen}}, \bibinfo
  {author} {\bibfnamefont {A.}~\bibnamefont {Rastelli}},\ and\ \bibinfo
  {author} {\bibfnamefont {S.}~\bibnamefont {Reitzenstein}},\ }\bibfield
  {title} {\bibinfo {title} {Quantum dots for photonic quantum information
  technology},\ }\href {https://doi.org/10.1364/AOP.490091} {\bibfield
  {journal} {\bibinfo  {journal} {Advances in Optics and Photonics}\ }\textbf
  {\bibinfo {volume} {15}},\ \bibinfo {pages} {613} (\bibinfo {year}
  {2023})}\BibitemShut {NoStop}%
\bibitem [{\citenamefont {O'brien}\ \emph {et~al.}(2009)\citenamefont
  {O'brien}, \citenamefont {Furusawa},\ and\ \citenamefont
  {Vu{\v{c}}kovi{\'c}}}]{o2009photonic}%
  \BibitemOpen
  \bibfield  {author} {\bibinfo {author} {\bibfnamefont {J.~L.}\ \bibnamefont
  {O'brien}}, \bibinfo {author} {\bibfnamefont {A.}~\bibnamefont {Furusawa}},\
  and\ \bibinfo {author} {\bibfnamefont {J.}~\bibnamefont
  {Vu{\v{c}}kovi{\'c}}},\ }\bibfield  {title} {\bibinfo {title} {Photonic
  quantum technologies},\ }\href {https://doi.org/10.1038/nphoton.2009.229}
  {\bibfield  {journal} {\bibinfo  {journal} {Nature photonics}\ }\textbf
  {\bibinfo {volume} {3}},\ \bibinfo {pages} {687} (\bibinfo {year}
  {2009})}\BibitemShut {NoStop}%
\bibitem [{\citenamefont {Uppu}\ \emph {et~al.}(2021)\citenamefont {Uppu},
  \citenamefont {Midolo}, \citenamefont {Zhou}, \citenamefont {Carolan},\ and\
  \citenamefont {Lodahl}}]{Uppu20211308}%
  \BibitemOpen
  \bibfield  {author} {\bibinfo {author} {\bibfnamefont {R.}~\bibnamefont
  {Uppu}}, \bibinfo {author} {\bibfnamefont {L.}~\bibnamefont {Midolo}},
  \bibinfo {author} {\bibfnamefont {X.}~\bibnamefont {Zhou}}, \bibinfo {author}
  {\bibfnamefont {J.}~\bibnamefont {Carolan}},\ and\ \bibinfo {author}
  {\bibfnamefont {P.}~\bibnamefont {Lodahl}},\ }\bibfield  {title} {\bibinfo
  {title} {Quantum-dot-based deterministic photon–emitter interfaces for
  scalable photonic quantum technology},\ }\href
  {https://doi.org/10.1038/s41565-021-00965-6} {\bibfield  {journal} {\bibinfo
  {journal} {Nature Nanotechnology}\ }\textbf {\bibinfo {volume} {16}},\
  \bibinfo {pages} {1308 – 1317} (\bibinfo {year} {2021})}\BibitemShut
  {NoStop}%
\bibitem [{\citenamefont {Duan}\ \emph {et~al.}(2001)\citenamefont {Duan},
  \citenamefont {Lukin}, \citenamefont {Cirac},\ and\ \citenamefont
  {Zoller}}]{Duan_2001}%
  \BibitemOpen
  \bibfield  {author} {\bibinfo {author} {\bibfnamefont {L.-M.}\ \bibnamefont
  {Duan}}, \bibinfo {author} {\bibfnamefont {M.~D.}\ \bibnamefont {Lukin}},
  \bibinfo {author} {\bibfnamefont {J.~I.}\ \bibnamefont {Cirac}},\ and\
  \bibinfo {author} {\bibfnamefont {P.}~\bibnamefont {Zoller}},\ }\bibfield
  {title} {\bibinfo {title} {Long-distance quantum communication with atomic
  ensembles and linear optics},\ }\href {https://doi.org/10.1038/35106500}
  {\bibfield  {journal} {\bibinfo  {journal} {Nature}\ }\textbf {\bibinfo
  {volume} {414}},\ \bibinfo {pages} {413–418} (\bibinfo {year}
  {2001})}\BibitemShut {NoStop}%
\bibitem [{\citenamefont {Lago-Rivera}\ \emph {et~al.}(2023)\citenamefont
  {Lago-Rivera}, \citenamefont {Rakonjac}, \citenamefont {Grandi},\ and\
  \citenamefont {Riedmatten}}]{Lago_Rivera_2023}%
  \BibitemOpen
  \bibfield  {author} {\bibinfo {author} {\bibfnamefont {D.}~\bibnamefont
  {Lago-Rivera}}, \bibinfo {author} {\bibfnamefont {J.~V.}\ \bibnamefont
  {Rakonjac}}, \bibinfo {author} {\bibfnamefont {S.}~\bibnamefont {Grandi}},\
  and\ \bibinfo {author} {\bibfnamefont {H.~d.}\ \bibnamefont {Riedmatten}},\
  }\bibfield  {title} {\bibinfo {title} {Long distance multiplexed quantum
  teleportation from a telecom photon to a solid-state qubit},\ }\href
  {https://doi.org/10.1038/s41467-023-37518-5} {\bibfield  {journal} {\bibinfo
  {journal} {Nature Communications}\ }\textbf {\bibinfo {volume} {14}},\
  \bibinfo {pages} {1889} (\bibinfo {year} {2023})}\BibitemShut {NoStop}%
\bibitem [{\citenamefont {Ladd}\ \emph {et~al.}(2010)\citenamefont {Ladd},
  \citenamefont {Jelezko}, \citenamefont {Laflamme}, \citenamefont {Nakamura},
  \citenamefont {Monroe},\ and\ \citenamefont {O’Brien}}]{Ladd_2010}%
  \BibitemOpen
  \bibfield  {author} {\bibinfo {author} {\bibfnamefont {T.~D.}\ \bibnamefont
  {Ladd}}, \bibinfo {author} {\bibfnamefont {F.}~\bibnamefont {Jelezko}},
  \bibinfo {author} {\bibfnamefont {R.}~\bibnamefont {Laflamme}}, \bibinfo
  {author} {\bibfnamefont {Y.}~\bibnamefont {Nakamura}}, \bibinfo {author}
  {\bibfnamefont {C.}~\bibnamefont {Monroe}},\ and\ \bibinfo {author}
  {\bibfnamefont {J.~L.}\ \bibnamefont {O’Brien}},\ }\bibfield  {title}
  {\bibinfo {title} {Quantum computers},\ }\href
  {https://doi.org/10.1038/nature08812} {\bibfield  {journal} {\bibinfo
  {journal} {Nature}\ }\textbf {\bibinfo {volume} {464}},\ \bibinfo {pages}
  {45–53} (\bibinfo {year} {2010})}\BibitemShut {NoStop}%
\bibitem [{\citenamefont {De~Raedt}\ \emph {et~al.}(2019)\citenamefont
  {De~Raedt}, \citenamefont {Jin}, \citenamefont {Willsch}, \citenamefont
  {Willsch}, \citenamefont {Yoshioka}, \citenamefont {Ito}, \citenamefont
  {Yuan},\ and\ \citenamefont {Michielsen}}]{De_Raedt_2019}%
  \BibitemOpen
  \bibfield  {author} {\bibinfo {author} {\bibfnamefont {H.}~\bibnamefont
  {De~Raedt}}, \bibinfo {author} {\bibfnamefont {F.}~\bibnamefont {Jin}},
  \bibinfo {author} {\bibfnamefont {D.}~\bibnamefont {Willsch}}, \bibinfo
  {author} {\bibfnamefont {M.}~\bibnamefont {Willsch}}, \bibinfo {author}
  {\bibfnamefont {N.}~\bibnamefont {Yoshioka}}, \bibinfo {author}
  {\bibfnamefont {N.}~\bibnamefont {Ito}}, \bibinfo {author} {\bibfnamefont
  {S.}~\bibnamefont {Yuan}},\ and\ \bibinfo {author} {\bibfnamefont
  {K.}~\bibnamefont {Michielsen}},\ }\bibfield  {title} {\bibinfo {title}
  {Massively parallel quantum computer simulator, eleven years later},\ }\href
  {https://doi.org/10.1016/j.cpc.2018.11.005} {\bibfield  {journal} {\bibinfo
  {journal} {Computer Physics Communications}\ }\textbf {\bibinfo {volume}
  {237}},\ \bibinfo {pages} {47–61} (\bibinfo {year} {2019})}\BibitemShut
  {NoStop}%
\bibitem [{\citenamefont {Aharonovich}\ \emph {et~al.}(2016)\citenamefont
  {Aharonovich}, \citenamefont {Englund},\ and\ \citenamefont
  {Toth}}]{aharonovich2016solid}%
  \BibitemOpen
  \bibfield  {author} {\bibinfo {author} {\bibfnamefont {I.}~\bibnamefont
  {Aharonovich}}, \bibinfo {author} {\bibfnamefont {D.}~\bibnamefont
  {Englund}},\ and\ \bibinfo {author} {\bibfnamefont {M.}~\bibnamefont
  {Toth}},\ }\bibfield  {title} {\bibinfo {title} {Solid-state single-photon
  emitters},\ }\href {https://doi.org/10.1038/nphoton.2016.186} {\bibfield
  {journal} {\bibinfo  {journal} {Nature photonics}\ }\textbf {\bibinfo
  {volume} {10}},\ \bibinfo {pages} {631} (\bibinfo {year} {2016})}\BibitemShut
  {NoStop}%
\bibitem [{\citenamefont {Heshami}\ \emph {et~al.}(2016)\citenamefont
  {Heshami}, \citenamefont {England}, \citenamefont {Humphreys}, \citenamefont
  {Bustard}, \citenamefont {Acosta}, \citenamefont {Nunn},\ and\ \citenamefont
  {Sussman}}]{heshami_2016}%
  \BibitemOpen
  \bibfield  {author} {\bibinfo {author} {\bibfnamefont {K.}~\bibnamefont
  {Heshami}}, \bibinfo {author} {\bibfnamefont {D.~G.}\ \bibnamefont
  {England}}, \bibinfo {author} {\bibfnamefont {P.~C.}\ \bibnamefont
  {Humphreys}}, \bibinfo {author} {\bibfnamefont {P.~J.}\ \bibnamefont
  {Bustard}}, \bibinfo {author} {\bibfnamefont {V.~M.}\ \bibnamefont {Acosta}},
  \bibinfo {author} {\bibfnamefont {J.}~\bibnamefont {Nunn}},\ and\ \bibinfo
  {author} {\bibfnamefont {B.~J.}\ \bibnamefont {Sussman}},\ }\bibfield
  {title} {\bibinfo {title} {Quantum memories: emerging applications and recent
  advances},\ }\href {https://doi.org/10.1080/09500340.2016.1148212} {\bibfield
   {journal} {\bibinfo  {journal} {Journal of modern optics}\ }\textbf
  {\bibinfo {volume} {63}},\ \bibinfo {pages} {2005} (\bibinfo {year}
  {2016})}\BibitemShut {NoStop}%
\bibitem [{\citenamefont {Kamel}\ \emph {et~al.}(2025)\citenamefont {Kamel},
  \citenamefont {Kumar}, \citenamefont {Gautam}, \citenamefont {Saglamyurek},
  \citenamefont {Salari},\ and\ \citenamefont {Oblak}}]{kamel-2025}%
  \BibitemOpen
  \bibfield  {author} {\bibinfo {author} {\bibfnamefont {N.~G.}\ \bibnamefont
  {Kamel}}, \bibinfo {author} {\bibfnamefont {S.}~\bibnamefont {Kumar}},
  \bibinfo {author} {\bibfnamefont {U.}~\bibnamefont {Gautam}}, \bibinfo
  {author} {\bibfnamefont {E.}~\bibnamefont {Saglamyurek}}, \bibinfo {author}
  {\bibfnamefont {V.}~\bibnamefont {Salari}},\ and\ \bibinfo {author}
  {\bibfnamefont {D.}~\bibnamefont {Oblak}},\ }\bibfield  {title} {\bibinfo
  {title} {{Multimode and Random-Access optical quantum memory via adiabatic
  phase imprinting}},\ }\bibfield  {journal} {\bibinfo  {journal} {arXiv
  preprint}\ }\href {https://doi.org/10.48550/arXiv.2506.12223}
  {10.48550/arXiv.2506.12223} (\bibinfo {year} {2025})\BibitemShut {NoStop}%
\bibitem [{\citenamefont {Thomas}\ \emph {et~al.}(2024)\citenamefont {Thomas},
  \citenamefont {Wagner}, \citenamefont {Joos}, \citenamefont {Sittig},
  \citenamefont {Nawrath}, \citenamefont {Burdekin}, \citenamefont
  {de~Buy~Wenniger}, \citenamefont {Rasiah}, \citenamefont {Huber-Loyola},
  \citenamefont {Sagona-Stophel} \emph {et~al.}}]{Thomas_2024}%
  \BibitemOpen
  \bibfield  {author} {\bibinfo {author} {\bibfnamefont {S.~E.}\ \bibnamefont
  {Thomas}}, \bibinfo {author} {\bibfnamefont {L.}~\bibnamefont {Wagner}},
  \bibinfo {author} {\bibfnamefont {R.}~\bibnamefont {Joos}}, \bibinfo {author}
  {\bibfnamefont {R.}~\bibnamefont {Sittig}}, \bibinfo {author} {\bibfnamefont
  {C.}~\bibnamefont {Nawrath}}, \bibinfo {author} {\bibfnamefont
  {P.}~\bibnamefont {Burdekin}}, \bibinfo {author} {\bibfnamefont {I.~M.}\
  \bibnamefont {de~Buy~Wenniger}}, \bibinfo {author} {\bibfnamefont {M.~J.}\
  \bibnamefont {Rasiah}}, \bibinfo {author} {\bibfnamefont {T.}~\bibnamefont
  {Huber-Loyola}}, \bibinfo {author} {\bibfnamefont {S.}~\bibnamefont
  {Sagona-Stophel}}, \emph {et~al.},\ }\bibfield  {title} {\bibinfo {title}
  {Deterministic storage and retrieval of telecom light from a quantum dot
  single-photon source interfaced with an atomic quantum memory},\ }\href
  {https://doi.org/10.1126/sciadv.adi7346} {\bibfield  {journal} {\bibinfo
  {journal} {Science Advances}\ }\textbf {\bibinfo {volume} {10}},\ \bibinfo
  {pages} {eadi7346} (\bibinfo {year} {2024})}\BibitemShut {NoStop}%
\bibitem [{\citenamefont {Maruf}\ \emph {et~al.}(2023)\citenamefont {Maruf},
  \citenamefont {Venuturumilli}, \citenamefont {Bharadwaj}, \citenamefont
  {Anderson}, \citenamefont {Qiu}, \citenamefont {Yuan}, \citenamefont
  {Zeeshan}, \citenamefont {Semnani}, \citenamefont {Poole}, \citenamefont
  {Dalacu} \emph {et~al.}}]{maruf_2023}%
  \BibitemOpen
  \bibfield  {author} {\bibinfo {author} {\bibfnamefont {R.~A.}\ \bibnamefont
  {Maruf}}, \bibinfo {author} {\bibfnamefont {S.}~\bibnamefont
  {Venuturumilli}}, \bibinfo {author} {\bibfnamefont {D.}~\bibnamefont
  {Bharadwaj}}, \bibinfo {author} {\bibfnamefont {P.}~\bibnamefont {Anderson}},
  \bibinfo {author} {\bibfnamefont {J.}~\bibnamefont {Qiu}}, \bibinfo {author}
  {\bibfnamefont {Y.}~\bibnamefont {Yuan}}, \bibinfo {author} {\bibfnamefont
  {M.}~\bibnamefont {Zeeshan}}, \bibinfo {author} {\bibfnamefont
  {B.}~\bibnamefont {Semnani}}, \bibinfo {author} {\bibfnamefont {P.~J.}\
  \bibnamefont {Poole}}, \bibinfo {author} {\bibfnamefont {D.}~\bibnamefont
  {Dalacu}}, \emph {et~al.},\ }\bibfield  {title} {\bibinfo {title} {Widely
  tunable solid-state source of single-photons matching an atomic transition},\
  }\href {https://doi.org/10.48550/arXiv.2309.06734} {\bibfield  {journal}
  {\bibinfo  {journal} {arXiv preprint arXiv:2309.06734}\ } (\bibinfo {year}
  {2023})}\BibitemShut {NoStop}%
\bibitem [{\citenamefont {Vural}\ \emph {et~al.}(2018)\citenamefont {Vural},
  \citenamefont {Portalupi}, \citenamefont {Maisch}, \citenamefont {Kern},
  \citenamefont {Weber}, \citenamefont {Jetter}, \citenamefont {Wrachtrup},
  \citenamefont {Löw}, \citenamefont {Gerhardt},\ and\ \citenamefont
  {Michler}}]{Vural_2018}%
  \BibitemOpen
  \bibfield  {author} {\bibinfo {author} {\bibfnamefont {H.}~\bibnamefont
  {Vural}}, \bibinfo {author} {\bibfnamefont {S.~L.}\ \bibnamefont
  {Portalupi}}, \bibinfo {author} {\bibfnamefont {J.}~\bibnamefont {Maisch}},
  \bibinfo {author} {\bibfnamefont {S.}~\bibnamefont {Kern}}, \bibinfo {author}
  {\bibfnamefont {J.~H.}\ \bibnamefont {Weber}}, \bibinfo {author}
  {\bibfnamefont {M.}~\bibnamefont {Jetter}}, \bibinfo {author} {\bibfnamefont
  {J.}~\bibnamefont {Wrachtrup}}, \bibinfo {author} {\bibfnamefont
  {R.}~\bibnamefont {Löw}}, \bibinfo {author} {\bibfnamefont {I.}~\bibnamefont
  {Gerhardt}},\ and\ \bibinfo {author} {\bibfnamefont {P.}~\bibnamefont
  {Michler}},\ }\bibfield  {title} {\bibinfo {title} {Two-photon interference
  in an atom–quantum dot hybrid system},\ }\href
  {https://doi.org/10.1364/optica.5.000367} {\bibfield  {journal} {\bibinfo
  {journal} {Optica}\ }\textbf {\bibinfo {volume} {5}},\ \bibinfo {pages} {367}
  (\bibinfo {year} {2018})}\BibitemShut {NoStop}%
\bibitem [{\citenamefont {Chen}\ \emph {et~al.}(2016)\citenamefont {Chen},
  \citenamefont {Zhang}, \citenamefont {Zopf}, \citenamefont {Jung},
  \citenamefont {Zhang}, \citenamefont {Keil}, \citenamefont {Ding},\ and\
  \citenamefont {Schmidt}}]{Chen_2016}%
  \BibitemOpen
  \bibfield  {author} {\bibinfo {author} {\bibfnamefont {Y.}~\bibnamefont
  {Chen}}, \bibinfo {author} {\bibfnamefont {J.}~\bibnamefont {Zhang}},
  \bibinfo {author} {\bibfnamefont {M.}~\bibnamefont {Zopf}}, \bibinfo {author}
  {\bibfnamefont {K.}~\bibnamefont {Jung}}, \bibinfo {author} {\bibfnamefont
  {Y.}~\bibnamefont {Zhang}}, \bibinfo {author} {\bibfnamefont
  {R.}~\bibnamefont {Keil}}, \bibinfo {author} {\bibfnamefont {F.}~\bibnamefont
  {Ding}},\ and\ \bibinfo {author} {\bibfnamefont {O.~G.}\ \bibnamefont
  {Schmidt}},\ }\bibfield  {title} {\bibinfo {title} {Wavelength-tunable
  entangled photons from silicon-integrated iii--v quantum dots},\ }\href
  {https://doi.org/10.1038/ncomms10387} {\bibfield  {journal} {\bibinfo
  {journal} {Nature communications}\ }\textbf {\bibinfo {volume} {7}},\
  \bibinfo {pages} {10387} (\bibinfo {year} {2016})}\BibitemShut {NoStop}%
\bibitem [{\citenamefont {Meyer}\ \emph {et~al.}(2015)\citenamefont {Meyer},
  \citenamefont {Stockill}, \citenamefont {Steiner}, \citenamefont {Le~Gall},
  \citenamefont {Matthiesen}, \citenamefont {Clarke}, \citenamefont {Ludwig},
  \citenamefont {Reichel}, \citenamefont {Atat{\"u}re},\ and\ \citenamefont
  {K{\"o}hl}}]{Meyer_2015}%
  \BibitemOpen
  \bibfield  {author} {\bibinfo {author} {\bibfnamefont {H.}~\bibnamefont
  {Meyer}}, \bibinfo {author} {\bibfnamefont {R.}~\bibnamefont {Stockill}},
  \bibinfo {author} {\bibfnamefont {M.}~\bibnamefont {Steiner}}, \bibinfo
  {author} {\bibfnamefont {C.}~\bibnamefont {Le~Gall}}, \bibinfo {author}
  {\bibfnamefont {C.}~\bibnamefont {Matthiesen}}, \bibinfo {author}
  {\bibfnamefont {E.}~\bibnamefont {Clarke}}, \bibinfo {author} {\bibfnamefont
  {A.}~\bibnamefont {Ludwig}}, \bibinfo {author} {\bibfnamefont
  {J.}~\bibnamefont {Reichel}}, \bibinfo {author} {\bibfnamefont
  {M.}~\bibnamefont {Atat{\"u}re}},\ and\ \bibinfo {author} {\bibfnamefont
  {M.}~\bibnamefont {K{\"o}hl}},\ }\bibfield  {title} {\bibinfo {title} {Direct
  photonic coupling of a semiconductor quantum dot and a trapped ion},\ }\href
  {https://doi.org/10.1103/physrevlett.114.123001} {\bibfield  {journal}
  {\bibinfo  {journal} {Physical review letters}\ }\textbf {\bibinfo {volume}
  {114}},\ \bibinfo {pages} {123001} (\bibinfo {year} {2015})}\BibitemShut
  {NoStop}%
\bibitem [{\citenamefont {Tang}\ \emph {et~al.}(2015)\citenamefont {Tang},
  \citenamefont {Zhou}, \citenamefont {Wang}, \citenamefont {Li}, \citenamefont
  {Liu}, \citenamefont {Hua}, \citenamefont {Zou}, \citenamefont {Wang},
  \citenamefont {He}, \citenamefont {Chen} \emph {et~al.}}]{Tang_2015}%
  \BibitemOpen
  \bibfield  {author} {\bibinfo {author} {\bibfnamefont {J.-S.}\ \bibnamefont
  {Tang}}, \bibinfo {author} {\bibfnamefont {Z.-Q.}\ \bibnamefont {Zhou}},
  \bibinfo {author} {\bibfnamefont {Y.-T.}\ \bibnamefont {Wang}}, \bibinfo
  {author} {\bibfnamefont {Y.-L.}\ \bibnamefont {Li}}, \bibinfo {author}
  {\bibfnamefont {X.}~\bibnamefont {Liu}}, \bibinfo {author} {\bibfnamefont
  {Y.-L.}\ \bibnamefont {Hua}}, \bibinfo {author} {\bibfnamefont
  {Y.}~\bibnamefont {Zou}}, \bibinfo {author} {\bibfnamefont {S.}~\bibnamefont
  {Wang}}, \bibinfo {author} {\bibfnamefont {D.-Y.}\ \bibnamefont {He}},
  \bibinfo {author} {\bibfnamefont {G.}~\bibnamefont {Chen}}, \emph {et~al.},\
  }\bibfield  {title} {\bibinfo {title} {Storage of multiple single-photon
  pulses emitted from a quantum dot in a solid-state quantum memory},\ }\href
  {https://doi.org/10.1038/ncomms9652} {\bibfield  {journal} {\bibinfo
  {journal} {Nature communications}\ }\textbf {\bibinfo {volume} {6}},\
  \bibinfo {pages} {8652} (\bibinfo {year} {2015})}\BibitemShut {NoStop}%
\bibitem [{\citenamefont {Akopian}\ \emph {et~al.}(2011)\citenamefont
  {Akopian}, \citenamefont {Wang}, \citenamefont {Rastelli}, \citenamefont
  {Schmidt},\ and\ \citenamefont {Zwiller}}]{Akopian_2011}%
  \BibitemOpen
  \bibfield  {author} {\bibinfo {author} {\bibfnamefont {N.}~\bibnamefont
  {Akopian}}, \bibinfo {author} {\bibfnamefont {L.}~\bibnamefont {Wang}},
  \bibinfo {author} {\bibfnamefont {A.}~\bibnamefont {Rastelli}}, \bibinfo
  {author} {\bibfnamefont {O.~G.}\ \bibnamefont {Schmidt}},\ and\ \bibinfo
  {author} {\bibfnamefont {V.}~\bibnamefont {Zwiller}},\ }\bibfield  {title}
  {\bibinfo {title} {Hybrid semiconductor-atomic interface: slowing down single
  photons from a quantum dot},\ }\href
  {https://doi.org/10.1038/nphoton.2011.16} {\bibfield  {journal} {\bibinfo
  {journal} {Nature Photonics}\ }\textbf {\bibinfo {volume} {5}},\ \bibinfo
  {pages} {230–233} (\bibinfo {year} {2011})}\BibitemShut {NoStop}%
\bibitem [{\citenamefont {Phoenix}\ \emph {et~al.}(2022)\citenamefont
  {Phoenix}, \citenamefont {Korkusinski}, \citenamefont {Dalacu}, \citenamefont
  {Poole}, \citenamefont {Zawadzki}, \citenamefont {Studenikin}, \citenamefont
  {Williams}, \citenamefont {Sachrajda},\ and\ \citenamefont
  {Gaudreau}}]{Phoenix_2022}%
  \BibitemOpen
  \bibfield  {author} {\bibinfo {author} {\bibfnamefont {J.}~\bibnamefont
  {Phoenix}}, \bibinfo {author} {\bibfnamefont {M.}~\bibnamefont
  {Korkusinski}}, \bibinfo {author} {\bibfnamefont {D.}~\bibnamefont {Dalacu}},
  \bibinfo {author} {\bibfnamefont {P.~J.}\ \bibnamefont {Poole}}, \bibinfo
  {author} {\bibfnamefont {P.}~\bibnamefont {Zawadzki}}, \bibinfo {author}
  {\bibfnamefont {S.}~\bibnamefont {Studenikin}}, \bibinfo {author}
  {\bibfnamefont {R.~L.}\ \bibnamefont {Williams}}, \bibinfo {author}
  {\bibfnamefont {A.~S.}\ \bibnamefont {Sachrajda}},\ and\ \bibinfo {author}
  {\bibfnamefont {L.}~\bibnamefont {Gaudreau}},\ }\bibfield  {title} {\bibinfo
  {title} {Magnetic tuning of tunnel coupling between \text{InAsP} double
  quantum dots in \text{InP} nanowires},\ }\href
  {https://doi.org/10.1038/s41598-022-08548-8} {\bibfield  {journal} {\bibinfo
  {journal} {Scientific Reports}\ }\textbf {\bibinfo {volume} {12}},\ \bibinfo
  {pages} {5100} (\bibinfo {year} {2022})}\BibitemShut {NoStop}%
\bibitem [{\citenamefont {Lee}\ \emph {et~al.}(2017)\citenamefont {Lee},
  \citenamefont {Murray}, \citenamefont {Bennett}, \citenamefont {Ellis},
  \citenamefont {Dangel}, \citenamefont {Farrer}, \citenamefont {Spencer},
  \citenamefont {Ritchie},\ and\ \citenamefont {Shields}}]{Lee_2017}%
  \BibitemOpen
  \bibfield  {author} {\bibinfo {author} {\bibfnamefont {J.}~\bibnamefont
  {Lee}}, \bibinfo {author} {\bibfnamefont {E.}~\bibnamefont {Murray}},
  \bibinfo {author} {\bibfnamefont {A.}~\bibnamefont {Bennett}}, \bibinfo
  {author} {\bibfnamefont {D.}~\bibnamefont {Ellis}}, \bibinfo {author}
  {\bibfnamefont {C.}~\bibnamefont {Dangel}}, \bibinfo {author} {\bibfnamefont
  {I.}~\bibnamefont {Farrer}}, \bibinfo {author} {\bibfnamefont
  {P.}~\bibnamefont {Spencer}}, \bibinfo {author} {\bibfnamefont {D.~A.}\
  \bibnamefont {Ritchie}},\ and\ \bibinfo {author} {\bibfnamefont
  {A.}~\bibnamefont {Shields}},\ }\bibfield  {title} {\bibinfo {title}
  {Electrically driven and electrically tunable quantum light sources},\
  }\bibfield  {journal} {\bibinfo  {journal} {Applied Physics Letters}\
  }\textbf {\bibinfo {volume} {110}},\ \href
  {https://doi.org/10.1063/1.4976197} {10.1063/1.4976197} (\bibinfo {year}
  {2017})\BibitemShut {NoStop}%
\bibitem [{\citenamefont {Neuwirth}\ \emph {et~al.}(2021)\citenamefont
  {Neuwirth}, \citenamefont {Basset}, \citenamefont {Rota}, \citenamefont
  {Roccia}, \citenamefont {Schimpf}, \citenamefont {J{\"o}ns}, \citenamefont
  {Rastelli},\ and\ \citenamefont {Trotta}}]{neuwirth2021quantum}%
  \BibitemOpen
  \bibfield  {author} {\bibinfo {author} {\bibfnamefont {J.}~\bibnamefont
  {Neuwirth}}, \bibinfo {author} {\bibfnamefont {F.~B.}\ \bibnamefont
  {Basset}}, \bibinfo {author} {\bibfnamefont {M.~B.}\ \bibnamefont {Rota}},
  \bibinfo {author} {\bibfnamefont {E.}~\bibnamefont {Roccia}}, \bibinfo
  {author} {\bibfnamefont {C.}~\bibnamefont {Schimpf}}, \bibinfo {author}
  {\bibfnamefont {K.~D.}\ \bibnamefont {J{\"o}ns}}, \bibinfo {author}
  {\bibfnamefont {A.}~\bibnamefont {Rastelli}},\ and\ \bibinfo {author}
  {\bibfnamefont {R.}~\bibnamefont {Trotta}},\ }\bibfield  {title} {\bibinfo
  {title} {Quantum dot technology for quantum repeaters: from entangled photon
  generation toward the integration with quantum memories},\ }\href
  {https://doi.org/10.1088/2633-4356/ac3d14} {\bibfield  {journal} {\bibinfo
  {journal} {Materials for Quantum Technology}\ }\textbf {\bibinfo {volume}
  {1}},\ \bibinfo {pages} {043001} (\bibinfo {year} {2021})}\BibitemShut
  {NoStop}%
\bibitem [{\citenamefont {Ruskuc}\ \emph {et~al.}(2024)\citenamefont {Ruskuc},
  \citenamefont {Wu}, \citenamefont {Green}, \citenamefont {Hermans},
  \citenamefont {Choi},\ and\ \citenamefont {Faraon}}]{ruskuc2024scalable}%
  \BibitemOpen
  \bibfield  {author} {\bibinfo {author} {\bibfnamefont {A.}~\bibnamefont
  {Ruskuc}}, \bibinfo {author} {\bibfnamefont {C.-J.}\ \bibnamefont {Wu}},
  \bibinfo {author} {\bibfnamefont {E.}~\bibnamefont {Green}}, \bibinfo
  {author} {\bibfnamefont {S.~L.}\ \bibnamefont {Hermans}}, \bibinfo {author}
  {\bibfnamefont {J.}~\bibnamefont {Choi}},\ and\ \bibinfo {author}
  {\bibfnamefont {A.}~\bibnamefont {Faraon}},\ }\bibfield  {title} {\bibinfo
  {title} {Scalable multipartite entanglement of remote rare-earth ion
  qubits},\ }\bibfield  {journal} {\bibinfo  {journal} {arXiv preprint}\ }\href
  {https://doi.org/10.48550/arXiv.2402.16224} {10.48550/arXiv.2402.16224}
  (\bibinfo {year} {2024})\BibitemShut {NoStop}%
\bibitem [{\citenamefont {Rodt}\ \emph {et~al.}(2020)\citenamefont {Rodt},
  \citenamefont {Reitzenstein},\ and\ \citenamefont {Heindel}}]{rodt_2020}%
  \BibitemOpen
  \bibfield  {author} {\bibinfo {author} {\bibfnamefont {S.}~\bibnamefont
  {Rodt}}, \bibinfo {author} {\bibfnamefont {S.}~\bibnamefont {Reitzenstein}},\
  and\ \bibinfo {author} {\bibfnamefont {T.}~\bibnamefont {Heindel}},\
  }\bibfield  {title} {\bibinfo {title} {Deterministically fabricated
  solid-state quantum-light sources},\ }\href
  {https://doi.org/10.1088/1361-648x/ab5e15} {\bibfield  {journal} {\bibinfo
  {journal} {Journal of Physics: Condensed Matter}\ }\textbf {\bibinfo {volume}
  {32}},\ \bibinfo {pages} {153003} (\bibinfo {year} {2020})}\BibitemShut
  {NoStop}%
\bibitem [{\citenamefont {Versteegh}\ \emph {et~al.}(2014)\citenamefont
  {Versteegh}, \citenamefont {Reimer}, \citenamefont {J{\"o}ns}, \citenamefont
  {Dalacu}, \citenamefont {Poole}, \citenamefont {Gulinatti}, \citenamefont
  {Giudice},\ and\ \citenamefont {Zwiller}}]{Versteegh_2014}%
  \BibitemOpen
  \bibfield  {author} {\bibinfo {author} {\bibfnamefont {M.~A.}\ \bibnamefont
  {Versteegh}}, \bibinfo {author} {\bibfnamefont {M.~E.}\ \bibnamefont
  {Reimer}}, \bibinfo {author} {\bibfnamefont {K.~D.}\ \bibnamefont
  {J{\"o}ns}}, \bibinfo {author} {\bibfnamefont {D.}~\bibnamefont {Dalacu}},
  \bibinfo {author} {\bibfnamefont {P.~J.}\ \bibnamefont {Poole}}, \bibinfo
  {author} {\bibfnamefont {A.}~\bibnamefont {Gulinatti}}, \bibinfo {author}
  {\bibfnamefont {A.}~\bibnamefont {Giudice}},\ and\ \bibinfo {author}
  {\bibfnamefont {V.}~\bibnamefont {Zwiller}},\ }\bibfield  {title} {\bibinfo
  {title} {Observation of strongly entangled photon pairs from a nanowire
  quantum dot},\ }\href {https://doi.org/10.1038/ncomms6298} {\bibfield
  {journal} {\bibinfo  {journal} {Nature communications}\ }\textbf {\bibinfo
  {volume} {5}},\ \bibinfo {pages} {5298} (\bibinfo {year} {2014})}\BibitemShut
  {NoStop}%
\bibitem [{\citenamefont {Laferri{\`e}re}\ \emph {et~al.}(2022)\citenamefont
  {Laferri{\`e}re}, \citenamefont {Yeung}, \citenamefont {Miron}, \citenamefont
  {Northeast}, \citenamefont {Haffouz}, \citenamefont {Lapointe}, \citenamefont
  {Korkusinski}, \citenamefont {Poole}, \citenamefont {Williams},\ and\
  \citenamefont {Dalacu}}]{Laferri_re_2022}%
  \BibitemOpen
  \bibfield  {author} {\bibinfo {author} {\bibfnamefont {P.}~\bibnamefont
  {Laferri{\`e}re}}, \bibinfo {author} {\bibfnamefont {E.}~\bibnamefont
  {Yeung}}, \bibinfo {author} {\bibfnamefont {I.}~\bibnamefont {Miron}},
  \bibinfo {author} {\bibfnamefont {D.~B.}\ \bibnamefont {Northeast}}, \bibinfo
  {author} {\bibfnamefont {S.}~\bibnamefont {Haffouz}}, \bibinfo {author}
  {\bibfnamefont {J.}~\bibnamefont {Lapointe}}, \bibinfo {author}
  {\bibfnamefont {M.}~\bibnamefont {Korkusinski}}, \bibinfo {author}
  {\bibfnamefont {P.~J.}\ \bibnamefont {Poole}}, \bibinfo {author}
  {\bibfnamefont {R.~L.}\ \bibnamefont {Williams}},\ and\ \bibinfo {author}
  {\bibfnamefont {D.}~\bibnamefont {Dalacu}},\ }\bibfield  {title} {\bibinfo
  {title} {Unity yield of deterministically positioned quantum dot single
  photon sources},\ }\href {https://doi.org/10.1038/s41598-022-10451-1}
  {\bibfield  {journal} {\bibinfo  {journal} {Scientific Reports}\ }\textbf
  {\bibinfo {volume} {12}},\ \bibinfo {pages} {6376} (\bibinfo {year}
  {2022})}\BibitemShut {NoStop}%
\bibitem [{\citenamefont {Yeung}\ \emph
  {et~al.}(2023{\natexlab{a}})\citenamefont {Yeung}, \citenamefont {Northeast},
  \citenamefont {Jin}, \citenamefont {Laferri{\`e}re}, \citenamefont
  {Korkusinski}, \citenamefont {Poole}, \citenamefont {Williams},\ and\
  \citenamefont {Dalacu}}]{yeung2023chip}%
  \BibitemOpen
  \bibfield  {author} {\bibinfo {author} {\bibfnamefont {E.}~\bibnamefont
  {Yeung}}, \bibinfo {author} {\bibfnamefont {D.~B.}\ \bibnamefont
  {Northeast}}, \bibinfo {author} {\bibfnamefont {J.}~\bibnamefont {Jin}},
  \bibinfo {author} {\bibfnamefont {P.}~\bibnamefont {Laferri{\`e}re}},
  \bibinfo {author} {\bibfnamefont {M.}~\bibnamefont {Korkusinski}}, \bibinfo
  {author} {\bibfnamefont {P.~J.}\ \bibnamefont {Poole}}, \bibinfo {author}
  {\bibfnamefont {R.~L.}\ \bibnamefont {Williams}},\ and\ \bibinfo {author}
  {\bibfnamefont {D.}~\bibnamefont {Dalacu}},\ }\bibfield  {title} {\bibinfo
  {title} {On-chip indistinguishable photons using iii-v nanowire/sin hybrid
  integration},\ }\href {https://doi.org/10.1103/PhysRevB.108.195417}
  {\bibfield  {journal} {\bibinfo  {journal} {Physical Review B}\ }\textbf
  {\bibinfo {volume} {108}},\ \bibinfo {pages} {195417} (\bibinfo {year}
  {2023}{\natexlab{a}})}\BibitemShut {NoStop}%
\bibitem [{\citenamefont {Wakileh}\ \emph {et~al.}(2024)\citenamefont
  {Wakileh}, \citenamefont {Yu}, \citenamefont {Dokuz}, \citenamefont
  {Haffouz}, \citenamefont {Wu}, \citenamefont {Lapointe}, \citenamefont
  {Northeast}, \citenamefont {Williams}, \citenamefont {Rotenberg},
  \citenamefont {Poole} \emph {et~al.}}]{wakileh2024single}%
  \BibitemOpen
  \bibfield  {author} {\bibinfo {author} {\bibfnamefont {A.~N.}\ \bibnamefont
  {Wakileh}}, \bibinfo {author} {\bibfnamefont {L.}~\bibnamefont {Yu}},
  \bibinfo {author} {\bibfnamefont {D.}~\bibnamefont {Dokuz}}, \bibinfo
  {author} {\bibfnamefont {S.}~\bibnamefont {Haffouz}}, \bibinfo {author}
  {\bibfnamefont {X.}~\bibnamefont {Wu}}, \bibinfo {author} {\bibfnamefont
  {J.}~\bibnamefont {Lapointe}}, \bibinfo {author} {\bibfnamefont {D.~B.}\
  \bibnamefont {Northeast}}, \bibinfo {author} {\bibfnamefont {R.~L.}\
  \bibnamefont {Williams}}, \bibinfo {author} {\bibfnamefont {N.}~\bibnamefont
  {Rotenberg}}, \bibinfo {author} {\bibfnamefont {P.~J.}\ \bibnamefont
  {Poole}}, \emph {et~al.},\ }\bibfield  {title} {\bibinfo {title} {Single
  photon emission in the telecom c-band from nanowire-based quantum dots},\
  }\bibfield  {journal} {\bibinfo  {journal} {Applied Physics Letters}\
  }\textbf {\bibinfo {volume} {124}},\ \href
  {https://doi.org/10.1063/5.0179234} {10.1063/5.0179234} (\bibinfo {year}
  {2024})\BibitemShut {NoStop}%
\bibitem [{\citenamefont {Haffouz}\ \emph {et~al.}(2020)\citenamefont
  {Haffouz}, \citenamefont {Poole}, \citenamefont {Jin}, \citenamefont {Wu},
  \citenamefont {Ginet}, \citenamefont {Mnaymneh}, \citenamefont {Dalacu},\
  and\ \citenamefont {Williams}}]{haffouz2020single}%
  \BibitemOpen
  \bibfield  {author} {\bibinfo {author} {\bibfnamefont {S.}~\bibnamefont
  {Haffouz}}, \bibinfo {author} {\bibfnamefont {P.}~\bibnamefont {Poole}},
  \bibinfo {author} {\bibfnamefont {J.}~\bibnamefont {Jin}}, \bibinfo {author}
  {\bibfnamefont {X.}~\bibnamefont {Wu}}, \bibinfo {author} {\bibfnamefont
  {L.}~\bibnamefont {Ginet}}, \bibinfo {author} {\bibfnamefont
  {K.}~\bibnamefont {Mnaymneh}}, \bibinfo {author} {\bibfnamefont
  {D.}~\bibnamefont {Dalacu}},\ and\ \bibinfo {author} {\bibfnamefont
  {R.}~\bibnamefont {Williams}},\ }\bibfield  {title} {\bibinfo {title} {Single
  quantum dot-in-a-rod embedded in a photonic nanowire waveguide for telecom
  band emission},\ }\bibfield  {journal} {\bibinfo  {journal} {Applied Physics
  Letters}\ }\textbf {\bibinfo {volume} {117}},\ \href
  {https://doi.org/10.1063/5.0020681} {10.1063/5.0020681} (\bibinfo {year}
  {2020})\BibitemShut {NoStop}%
\bibitem [{\citenamefont {Carnall}\ \emph {et~al.}(1989)\citenamefont
  {Carnall}, \citenamefont {Goodman}, \citenamefont {Rajnak},\ and\
  \citenamefont {Rana}}]{carnall1989systematic}%
  \BibitemOpen
  \bibfield  {author} {\bibinfo {author} {\bibfnamefont {W.~T.}\ \bibnamefont
  {Carnall}}, \bibinfo {author} {\bibfnamefont {G.~L.}\ \bibnamefont
  {Goodman}}, \bibinfo {author} {\bibfnamefont {K.}~\bibnamefont {Rajnak}},\
  and\ \bibinfo {author} {\bibfnamefont {R.~S.}\ \bibnamefont {Rana}},\
  }\bibfield  {title} {\bibinfo {title} {A systematic analysis of the spectra
  of the lanthanides doped into single crystal laf$_3$},\ }\href
  {https://doi.org/10.1063/1.455853} {\bibfield  {journal} {\bibinfo  {journal}
  {The Journal of chemical physics}\ }\textbf {\bibinfo {volume} {90}},\
  \bibinfo {pages} {3443} (\bibinfo {year} {1989})}\BibitemShut {NoStop}%
\bibitem [{\citenamefont {Tanabe}(1999)}]{tanabe1999optical}%
  \BibitemOpen
  \bibfield  {author} {\bibinfo {author} {\bibfnamefont {S.}~\bibnamefont
  {Tanabe}},\ }\bibfield  {title} {\bibinfo {title} {Optical transitions of
  rare earth ions for amplifiers: how the local structure works in glass},\
  }\href {https://doi.org/10.1016/S0022-3093(99)00490-1} {\bibfield  {journal}
  {\bibinfo  {journal} {Journal of non-crystalline solids}\ }\textbf {\bibinfo
  {volume} {259}},\ \bibinfo {pages} {1} (\bibinfo {year} {1999})}\BibitemShut
  {NoStop}%
\bibitem [{\citenamefont {Zhong}\ \emph {et~al.}(2015)\citenamefont {Zhong},
  \citenamefont {Hedges}, \citenamefont {Ahlefeldt}, \citenamefont
  {Bartholomew}, \citenamefont {Beavan}, \citenamefont {Wittig}, \citenamefont
  {Longdell},\ and\ \citenamefont {Sellars}}]{Zhong_2015}%
  \BibitemOpen
  \bibfield  {author} {\bibinfo {author} {\bibfnamefont {M.}~\bibnamefont
  {Zhong}}, \bibinfo {author} {\bibfnamefont {M.~P.}\ \bibnamefont {Hedges}},
  \bibinfo {author} {\bibfnamefont {R.~L.}\ \bibnamefont {Ahlefeldt}}, \bibinfo
  {author} {\bibfnamefont {J.~G.}\ \bibnamefont {Bartholomew}}, \bibinfo
  {author} {\bibfnamefont {S.~E.}\ \bibnamefont {Beavan}}, \bibinfo {author}
  {\bibfnamefont {S.~M.}\ \bibnamefont {Wittig}}, \bibinfo {author}
  {\bibfnamefont {J.~J.}\ \bibnamefont {Longdell}},\ and\ \bibinfo {author}
  {\bibfnamefont {M.~J.}\ \bibnamefont {Sellars}},\ }\bibfield  {title}
  {\bibinfo {title} {Optically addressable nuclear spins in a solid with a
  six-hour coherence time},\ }\href {https://doi.org/10.1038/nature14025}
  {\bibfield  {journal} {\bibinfo  {journal} {Nature}\ }\textbf {\bibinfo
  {volume} {517}},\ \bibinfo {pages} {177–180} (\bibinfo {year}
  {2015})}\BibitemShut {NoStop}%
\bibitem [{\citenamefont {Nicolas}\ \emph {et~al.}(2023)\citenamefont
  {Nicolas}, \citenamefont {Businger}, \citenamefont {Sanchez~Mejia},
  \citenamefont {Tiranov}, \citenamefont {Chaneli{\`e}re}, \citenamefont
  {Lafitte-Houssat}, \citenamefont {Ferrier}, \citenamefont {Goldner},\ and\
  \citenamefont {Afzelius}}]{Nicolas_2023}%
  \BibitemOpen
  \bibfield  {author} {\bibinfo {author} {\bibfnamefont {L.}~\bibnamefont
  {Nicolas}}, \bibinfo {author} {\bibfnamefont {M.}~\bibnamefont {Businger}},
  \bibinfo {author} {\bibfnamefont {T.}~\bibnamefont {Sanchez~Mejia}}, \bibinfo
  {author} {\bibfnamefont {A.}~\bibnamefont {Tiranov}}, \bibinfo {author}
  {\bibfnamefont {T.}~\bibnamefont {Chaneli{\`e}re}}, \bibinfo {author}
  {\bibfnamefont {E.}~\bibnamefont {Lafitte-Houssat}}, \bibinfo {author}
  {\bibfnamefont {A.}~\bibnamefont {Ferrier}}, \bibinfo {author} {\bibfnamefont
  {P.}~\bibnamefont {Goldner}},\ and\ \bibinfo {author} {\bibfnamefont
  {M.}~\bibnamefont {Afzelius}},\ }\bibfield  {title} {\bibinfo {title}
  {Coherent optical-microwave interface for manipulation of low-field
  electronic clock transitions in
  $^{171}\mathrm{Yb}^{3+}:\mathrm{Y}_{2}\mathrm{SiO}_{5}$},\ }\href
  {https://doi.org/10.1038/s41534-023-00687-8} {\bibfield  {journal} {\bibinfo
  {journal} {npj Quantum Information}\ }\textbf {\bibinfo {volume} {9}},\
  \bibinfo {pages} {21} (\bibinfo {year} {2023})}\BibitemShut {NoStop}%
\bibitem [{\citenamefont {Wei}\ \emph {et~al.}(2024)\citenamefont {Wei},
  \citenamefont {Jing}, \citenamefont {Zhang}, \citenamefont {Liao},
  \citenamefont {Li}, \citenamefont {You}, \citenamefont {Wang}, \citenamefont
  {Wang}, \citenamefont {Deng}, \citenamefont {Song} \emph
  {et~al.}}]{Wei_2024}%
  \BibitemOpen
  \bibfield  {author} {\bibinfo {author} {\bibfnamefont {S.-H.}\ \bibnamefont
  {Wei}}, \bibinfo {author} {\bibfnamefont {B.}~\bibnamefont {Jing}}, \bibinfo
  {author} {\bibfnamefont {X.-Y.}\ \bibnamefont {Zhang}}, \bibinfo {author}
  {\bibfnamefont {J.-Y.}\ \bibnamefont {Liao}}, \bibinfo {author}
  {\bibfnamefont {H.}~\bibnamefont {Li}}, \bibinfo {author} {\bibfnamefont
  {L.-X.}\ \bibnamefont {You}}, \bibinfo {author} {\bibfnamefont
  {Z.}~\bibnamefont {Wang}}, \bibinfo {author} {\bibfnamefont {Y.}~\bibnamefont
  {Wang}}, \bibinfo {author} {\bibfnamefont {G.-W.}\ \bibnamefont {Deng}},
  \bibinfo {author} {\bibfnamefont {H.-Z.}\ \bibnamefont {Song}}, \emph
  {et~al.},\ }\bibfield  {title} {\bibinfo {title} {Quantum storage of 1650
  modes of single photons at telecom wavelength},\ }\href
  {https://doi.org/10.1038/s41534-024-00812-1} {\bibfield  {journal} {\bibinfo
  {journal} {npj Quantum Information}\ }\textbf {\bibinfo {volume} {10}},\
  \bibinfo {pages} {19} (\bibinfo {year} {2024})}\BibitemShut {NoStop}%
\bibitem [{\citenamefont {Businger}\ \emph {et~al.}(2022)\citenamefont
  {Businger}, \citenamefont {Nicolas}, \citenamefont {Mejia}, \citenamefont
  {Ferrier}, \citenamefont {Goldner},\ and\ \citenamefont
  {Afzelius}}]{Businger_2022}%
  \BibitemOpen
  \bibfield  {author} {\bibinfo {author} {\bibfnamefont {M.}~\bibnamefont
  {Businger}}, \bibinfo {author} {\bibfnamefont {L.}~\bibnamefont {Nicolas}},
  \bibinfo {author} {\bibfnamefont {T.~S.}\ \bibnamefont {Mejia}}, \bibinfo
  {author} {\bibfnamefont {A.}~\bibnamefont {Ferrier}}, \bibinfo {author}
  {\bibfnamefont {P.}~\bibnamefont {Goldner}},\ and\ \bibinfo {author}
  {\bibfnamefont {M.}~\bibnamefont {Afzelius}},\ }\bibfield  {title} {\bibinfo
  {title} {Non-classical correlations over 1250 modes between telecom photons
  and 979-nm photons stored in
  $^{171}\mathrm{Yb}^{3+}:\mathrm{Y}_{2}\mathrm{SiO}_{5}$},\ }\href
  {https://doi.org/10.1038/s41467-022-33929-y} {\bibfield  {journal} {\bibinfo
  {journal} {Nature communications}\ }\textbf {\bibinfo {volume} {13}},\
  \bibinfo {pages} {6438} (\bibinfo {year} {2022})}\BibitemShut {NoStop}%
\bibitem [{\citenamefont {Ma}\ \emph {et~al.}(2021)\citenamefont {Ma},
  \citenamefont {Ma}, \citenamefont {Zhou}, \citenamefont {Li},\ and\
  \citenamefont {Guo}}]{Ma_2021}%
  \BibitemOpen
  \bibfield  {author} {\bibinfo {author} {\bibfnamefont {Y.}~\bibnamefont
  {Ma}}, \bibinfo {author} {\bibfnamefont {Y.-Z.}\ \bibnamefont {Ma}}, \bibinfo
  {author} {\bibfnamefont {Z.-Q.}\ \bibnamefont {Zhou}}, \bibinfo {author}
  {\bibfnamefont {C.-F.}\ \bibnamefont {Li}},\ and\ \bibinfo {author}
  {\bibfnamefont {G.-C.}\ \bibnamefont {Guo}},\ }\bibfield  {title} {\bibinfo
  {title} {One-hour coherent optical storage in an atomic frequency comb
  memory},\ }\href {https://doi.org/10.1038/s41467-021-22706-y} {\bibfield
  {journal} {\bibinfo  {journal} {Nature communications}\ }\textbf {\bibinfo
  {volume} {12}},\ \bibinfo {pages} {2381} (\bibinfo {year}
  {2021})}\BibitemShut {NoStop}%
\bibitem [{\citenamefont {Ortu}\ \emph {et~al.}(2022)\citenamefont {Ortu},
  \citenamefont {Rakonjac}, \citenamefont {Holzäpfel}, \citenamefont {Seri},
  \citenamefont {Grandi}, \citenamefont {Mazzera}, \citenamefont
  {de~Riedmatten},\ and\ \citenamefont {Afzelius}}]{Ortu_2022}%
  \BibitemOpen
  \bibfield  {author} {\bibinfo {author} {\bibfnamefont {A.}~\bibnamefont
  {Ortu}}, \bibinfo {author} {\bibfnamefont {J.~V.}\ \bibnamefont {Rakonjac}},
  \bibinfo {author} {\bibfnamefont {A.}~\bibnamefont {Holzäpfel}}, \bibinfo
  {author} {\bibfnamefont {A.}~\bibnamefont {Seri}}, \bibinfo {author}
  {\bibfnamefont {S.}~\bibnamefont {Grandi}}, \bibinfo {author} {\bibfnamefont
  {M.}~\bibnamefont {Mazzera}}, \bibinfo {author} {\bibfnamefont
  {H.}~\bibnamefont {de~Riedmatten}},\ and\ \bibinfo {author} {\bibfnamefont
  {M.}~\bibnamefont {Afzelius}},\ }\bibfield  {title} {\bibinfo {title}
  {Multimode capacity of atomic-frequency comb quantum memories},\ }\href
  {https://doi.org/10.1088/2058-9565/ac73b0} {\bibfield  {journal} {\bibinfo
  {journal} {Quantum Science and Technology}\ }\textbf {\bibinfo {volume}
  {7}},\ \bibinfo {pages} {035024} (\bibinfo {year} {2022})}\BibitemShut
  {NoStop}%
\bibitem [{\citenamefont {Saglamyurek}\ \emph
  {et~al.}(2015{\natexlab{a}})\citenamefont {Saglamyurek}, \citenamefont {Jin},
  \citenamefont {Verma}, \citenamefont {Shaw}, \citenamefont {Marsili},
  \citenamefont {Nam}, \citenamefont {Oblak},\ and\ \citenamefont
  {Tittel}}]{Saglamyurek_2015}%
  \BibitemOpen
  \bibfield  {author} {\bibinfo {author} {\bibfnamefont {E.}~\bibnamefont
  {Saglamyurek}}, \bibinfo {author} {\bibfnamefont {J.}~\bibnamefont {Jin}},
  \bibinfo {author} {\bibfnamefont {V.~B.}\ \bibnamefont {Verma}}, \bibinfo
  {author} {\bibfnamefont {M.~D.}\ \bibnamefont {Shaw}}, \bibinfo {author}
  {\bibfnamefont {F.}~\bibnamefont {Marsili}}, \bibinfo {author} {\bibfnamefont
  {S.~W.}\ \bibnamefont {Nam}}, \bibinfo {author} {\bibfnamefont
  {D.}~\bibnamefont {Oblak}},\ and\ \bibinfo {author} {\bibfnamefont
  {W.}~\bibnamefont {Tittel}},\ }\bibfield  {title} {\bibinfo {title} {Quantum
  storage of entangled telecom-wavelength photons in an erbium-doped optical
  fibre},\ }\href {https://doi.org/10.1038/nphoton.2014.311} {\bibfield
  {journal} {\bibinfo  {journal} {Nature Photonics}\ }\textbf {\bibinfo
  {volume} {9}},\ \bibinfo {pages} {83–87} (\bibinfo {year}
  {2015}{\natexlab{a}})}\BibitemShut {NoStop}%
\bibitem [{\citenamefont {Jiang}\ \emph {et~al.}(2023)\citenamefont {Jiang},
  \citenamefont {Xue}, \citenamefont {He}, \citenamefont {An}, \citenamefont
  {Zheng}, \citenamefont {Xu}, \citenamefont {Xie}, \citenamefont {Lu},
  \citenamefont {Zhu},\ and\ \citenamefont {Ma}}]{jiang2023quantum}%
  \BibitemOpen
  \bibfield  {author} {\bibinfo {author} {\bibfnamefont {M.-H.}\ \bibnamefont
  {Jiang}}, \bibinfo {author} {\bibfnamefont {W.}~\bibnamefont {Xue}}, \bibinfo
  {author} {\bibfnamefont {Q.}~\bibnamefont {He}}, \bibinfo {author}
  {\bibfnamefont {Y.-Y.}\ \bibnamefont {An}}, \bibinfo {author} {\bibfnamefont
  {X.}~\bibnamefont {Zheng}}, \bibinfo {author} {\bibfnamefont {W.-J.}\
  \bibnamefont {Xu}}, \bibinfo {author} {\bibfnamefont {Y.-B.}\ \bibnamefont
  {Xie}}, \bibinfo {author} {\bibfnamefont {Y.}~\bibnamefont {Lu}}, \bibinfo
  {author} {\bibfnamefont {S.}~\bibnamefont {Zhu}},\ and\ \bibinfo {author}
  {\bibfnamefont {X.-S.}\ \bibnamefont {Ma}},\ }\bibfield  {title} {\bibinfo
  {title} {Quantum storage of entangled photons at telecom wavelengths in a
  crystal},\ }\href {https://doi.org/10.1038/s41467-023-42741-1} {\bibfield
  {journal} {\bibinfo  {journal} {Nature Communications}\ }\textbf {\bibinfo
  {volume} {14}},\ \bibinfo {pages} {6995} (\bibinfo {year}
  {2023})}\BibitemShut {NoStop}%
\bibitem [{\citenamefont {Lago-Rivera}\ \emph {et~al.}(2021)\citenamefont
  {Lago-Rivera}, \citenamefont {Grandi}, \citenamefont {Rakonjac},
  \citenamefont {Seri},\ and\ \citenamefont
  {de~Riedmatten}}]{Lago-Rivera202137}%
  \BibitemOpen
  \bibfield  {author} {\bibinfo {author} {\bibfnamefont {D.}~\bibnamefont
  {Lago-Rivera}}, \bibinfo {author} {\bibfnamefont {S.}~\bibnamefont {Grandi}},
  \bibinfo {author} {\bibfnamefont {J.~V.}\ \bibnamefont {Rakonjac}}, \bibinfo
  {author} {\bibfnamefont {A.}~\bibnamefont {Seri}},\ and\ \bibinfo {author}
  {\bibfnamefont {H.}~\bibnamefont {de~Riedmatten}},\ }\bibfield  {title}
  {\bibinfo {title} {Telecom-heralded entanglement between multimode
  solid-state quantum memories},\ }\href
  {https://doi.org/10.1038/s41586-021-03481-8} {\bibfield  {journal} {\bibinfo
  {journal} {Nature}\ }\textbf {\bibinfo {volume} {594}},\ \bibinfo {pages} {37
  – 40} (\bibinfo {year} {2021})}\BibitemShut {NoStop}%
\bibitem [{\citenamefont {Puigibert}\ \emph {et~al.}(2020)\citenamefont
  {Puigibert}, \citenamefont {Askarani}, \citenamefont {Davidson},
  \citenamefont {Verma}, \citenamefont {Shaw}, \citenamefont {Nam},
  \citenamefont {Lutz}, \citenamefont {Amaral}, \citenamefont {Oblak},\ and\
  \citenamefont {Tittel}}]{puigibert2020entanglement}%
  \BibitemOpen
  \bibfield  {author} {\bibinfo {author} {\bibfnamefont {M.~L.~G.}\
  \bibnamefont {Puigibert}}, \bibinfo {author} {\bibfnamefont {M.~F.}\
  \bibnamefont {Askarani}}, \bibinfo {author} {\bibfnamefont {J.~H.}\
  \bibnamefont {Davidson}}, \bibinfo {author} {\bibfnamefont {V.~B.}\
  \bibnamefont {Verma}}, \bibinfo {author} {\bibfnamefont {M.~D.}\ \bibnamefont
  {Shaw}}, \bibinfo {author} {\bibfnamefont {S.~W.}\ \bibnamefont {Nam}},
  \bibinfo {author} {\bibfnamefont {T.}~\bibnamefont {Lutz}}, \bibinfo {author}
  {\bibfnamefont {G.~C.}\ \bibnamefont {Amaral}}, \bibinfo {author}
  {\bibfnamefont {D.}~\bibnamefont {Oblak}},\ and\ \bibinfo {author}
  {\bibfnamefont {W.}~\bibnamefont {Tittel}},\ }\bibfield  {title} {\bibinfo
  {title} {Entanglement and nonlocality between disparate solid-state quantum
  memories mediated by photons},\ }\href
  {https://doi.org/10.1103/PhysRevResearch.2.013039} {\bibfield  {journal}
  {\bibinfo  {journal} {Physical Review Research}\ }\textbf {\bibinfo {volume}
  {2}},\ \bibinfo {pages} {013039} (\bibinfo {year} {2020})}\BibitemShut
  {NoStop}%
\bibitem [{\citenamefont {Aharonovich}\ and\ \citenamefont
  {Neu}(2014)}]{aharonovich2014diamond}%
  \BibitemOpen
  \bibfield  {author} {\bibinfo {author} {\bibfnamefont {I.}~\bibnamefont
  {Aharonovich}}\ and\ \bibinfo {author} {\bibfnamefont {E.}~\bibnamefont
  {Neu}},\ }\bibfield  {title} {\bibinfo {title} {Diamond nanophotonics},\
  }\href {https://doi.org/10.1002/adom.201400189} {\bibfield  {journal}
  {\bibinfo  {journal} {Advanced Optical Materials}\ }\textbf {\bibinfo
  {volume} {2}},\ \bibinfo {pages} {911} (\bibinfo {year} {2014})}\BibitemShut
  {NoStop}%
\bibitem [{\citenamefont {Li}\ \emph {et~al.}(2023)\citenamefont {Li},
  \citenamefont {Liu},\ and\ \citenamefont {Lu}}]{Li2023}%
  \BibitemOpen
  \bibfield  {author} {\bibinfo {author} {\bibfnamefont {R.}~\bibnamefont
  {Li}}, \bibinfo {author} {\bibfnamefont {F.}~\bibnamefont {Liu}},\ and\
  \bibinfo {author} {\bibfnamefont {Q.}~\bibnamefont {Lu}},\ }\bibfield
  {title} {\bibinfo {title} {Quantum light source based on semiconductor
  quantum dots: a review},\ }\href {https://doi.org/10.3390/photonics10060639}
  {\bibfield  {journal} {\bibinfo  {journal} {Photonics}\ }\textbf {\bibinfo
  {volume} {10}},\ \bibinfo {pages} {639} (\bibinfo {year} {2023})}\BibitemShut
  {NoStop}%
\bibitem [{\citenamefont {Garc{\'\i}a~de Arquer}\ \emph
  {et~al.}(2021)\citenamefont {Garc{\'\i}a~de Arquer}, \citenamefont {Talapin},
  \citenamefont {Klimov}, \citenamefont {Arakawa}, \citenamefont {Bayer},\ and\
  \citenamefont {Sargent}}]{garcia2021semiconductor}%
  \BibitemOpen
  \bibfield  {author} {\bibinfo {author} {\bibfnamefont {F.~P.}\ \bibnamefont
  {Garc{\'\i}a~de Arquer}}, \bibinfo {author} {\bibfnamefont {D.~V.}\
  \bibnamefont {Talapin}}, \bibinfo {author} {\bibfnamefont {V.~I.}\
  \bibnamefont {Klimov}}, \bibinfo {author} {\bibfnamefont {Y.}~\bibnamefont
  {Arakawa}}, \bibinfo {author} {\bibfnamefont {M.}~\bibnamefont {Bayer}},\
  and\ \bibinfo {author} {\bibfnamefont {E.~H.}\ \bibnamefont {Sargent}},\
  }\bibfield  {title} {\bibinfo {title} {Semiconductor quantum dots:
  Technological progress and future challenges},\ }\href
  {https://doi.org/10.1126/science.aaz854} {\bibfield  {journal} {\bibinfo
  {journal} {Science}\ }\textbf {\bibinfo {volume} {373}},\ \bibinfo {pages}
  {eaaz8541} (\bibinfo {year} {2021})}\BibitemShut {NoStop}%
\bibitem [{\citenamefont {Couteau}(2018)}]{Couteau2018291}%
  \BibitemOpen
  \bibfield  {author} {\bibinfo {author} {\bibfnamefont {C.}~\bibnamefont
  {Couteau}},\ }\bibfield  {title} {\bibinfo {title} {Spontaneous parametric
  down-conversion},\ }\href {https://doi.org/10.1080/00107514.2018.1488463}
  {\bibfield  {journal} {\bibinfo  {journal} {Contemporary Physics}\ }\textbf
  {\bibinfo {volume} {59}},\ \bibinfo {pages} {291 – 304} (\bibinfo {year}
  {2018})}\BibitemShut {NoStop}%
\bibitem [{\citenamefont {Afzelius}\ \emph {et~al.}(2009)\citenamefont
  {Afzelius}, \citenamefont {Simon}, \citenamefont {De~Riedmatten},\ and\
  \citenamefont {Gisin}}]{Afzelius_2009}%
  \BibitemOpen
  \bibfield  {author} {\bibinfo {author} {\bibfnamefont {M.}~\bibnamefont
  {Afzelius}}, \bibinfo {author} {\bibfnamefont {C.}~\bibnamefont {Simon}},
  \bibinfo {author} {\bibfnamefont {H.}~\bibnamefont {De~Riedmatten}},\ and\
  \bibinfo {author} {\bibfnamefont {N.}~\bibnamefont {Gisin}},\ }\bibfield
  {title} {\bibinfo {title} {Multimode quantum memory based on atomic frequency
  combs},\ }\href {https://doi.org/10.1103/physreva.79.052329} {\bibfield
  {journal} {\bibinfo  {journal} {Physical Review A—Atomic, Molecular, and
  Optical Physics}\ }\textbf {\bibinfo {volume} {79}},\ \bibinfo {pages}
  {052329} (\bibinfo {year} {2009})}\BibitemShut {NoStop}%
\bibitem [{\citenamefont {Saglamyurek}\ \emph {et~al.}(2011)\citenamefont
  {Saglamyurek}, \citenamefont {Sinclair}, \citenamefont {Jin}, \citenamefont
  {Slater}, \citenamefont {Oblak}, \citenamefont {Bussieres}, \citenamefont
  {George}, \citenamefont {Ricken}, \citenamefont {Sohler},\ and\ \citenamefont
  {Tittel}}]{saglamyurek2011broadband}%
  \BibitemOpen
  \bibfield  {author} {\bibinfo {author} {\bibfnamefont {E.}~\bibnamefont
  {Saglamyurek}}, \bibinfo {author} {\bibfnamefont {N.}~\bibnamefont
  {Sinclair}}, \bibinfo {author} {\bibfnamefont {J.}~\bibnamefont {Jin}},
  \bibinfo {author} {\bibfnamefont {J.~A.}\ \bibnamefont {Slater}}, \bibinfo
  {author} {\bibfnamefont {D.}~\bibnamefont {Oblak}}, \bibinfo {author}
  {\bibfnamefont {F.}~\bibnamefont {Bussieres}}, \bibinfo {author}
  {\bibfnamefont {M.}~\bibnamefont {George}}, \bibinfo {author} {\bibfnamefont
  {R.}~\bibnamefont {Ricken}}, \bibinfo {author} {\bibfnamefont
  {W.}~\bibnamefont {Sohler}},\ and\ \bibinfo {author} {\bibfnamefont
  {W.}~\bibnamefont {Tittel}},\ }\bibfield  {title} {\bibinfo {title}
  {Broadband waveguide quantum memory for entangled photons},\ }\href
  {https://doi.org/10.1038/nature09719} {\bibfield  {journal} {\bibinfo
  {journal} {Nature}\ }\textbf {\bibinfo {volume} {469}},\ \bibinfo {pages}
  {512} (\bibinfo {year} {2011})}\BibitemShut {NoStop}%
\bibitem [{\citenamefont {Brida}\ \emph {et~al.}(2014)\citenamefont {Brida},
  \citenamefont {Krauss}, \citenamefont {Sell},\ and\ \citenamefont
  {Leitenstorfer}}]{Brida2014409}%
  \BibitemOpen
  \bibfield  {author} {\bibinfo {author} {\bibfnamefont {D.}~\bibnamefont
  {Brida}}, \bibinfo {author} {\bibfnamefont {G.}~\bibnamefont {Krauss}},
  \bibinfo {author} {\bibfnamefont {A.}~\bibnamefont {Sell}},\ and\ \bibinfo
  {author} {\bibfnamefont {A.}~\bibnamefont {Leitenstorfer}},\ }\bibfield
  {title} {\bibinfo {title} {Ultrabroadband er: Fiber lasers},\ }\href
  {https://doi.org/10.1002/lpor.201300194} {\bibfield  {journal} {\bibinfo
  {journal} {Laser and Photonics Reviews}\ }\textbf {\bibinfo {volume} {8}},\
  \bibinfo {pages} {409 – 428} (\bibinfo {year} {2014})}\BibitemShut
  {NoStop}%
\bibitem [{See the Supplementary Information()}]{Kamel_spp}%
  \BibitemOpen
  See the Supplementary Information,\ \href@noop {} {}\BibitemShut {NoStop}%
\bibitem [{\citenamefont {Stachurski}\ \emph {et~al.}(2022)\citenamefont
  {Stachurski}, \citenamefont {Tamariz}, \citenamefont {Callsen}, \citenamefont
  {Butt{\'e}},\ and\ \citenamefont {Grandjean}}]{stachurski2022single}%
  \BibitemOpen
  \bibfield  {author} {\bibinfo {author} {\bibfnamefont {J.}~\bibnamefont
  {Stachurski}}, \bibinfo {author} {\bibfnamefont {S.}~\bibnamefont {Tamariz}},
  \bibinfo {author} {\bibfnamefont {G.}~\bibnamefont {Callsen}}, \bibinfo
  {author} {\bibfnamefont {R.}~\bibnamefont {Butt{\'e}}},\ and\ \bibinfo
  {author} {\bibfnamefont {N.}~\bibnamefont {Grandjean}},\ }\bibfield  {title}
  {\bibinfo {title} {Single photon emission and recombination dynamics in
  self-assembled gan/aln quantum dots},\ }\href
  {https://doi.org/10.1038/s41377-022-00799-4} {\bibfield  {journal} {\bibinfo
  {journal} {Light: Science \& Applications}\ }\textbf {\bibinfo {volume}
  {11}},\ \bibinfo {pages} {114} (\bibinfo {year} {2022})}\BibitemShut
  {NoStop}%
\bibitem [{\citenamefont {Holewa}\ \emph {et~al.}(2024)\citenamefont {Holewa},
  \citenamefont {Vajner}, \citenamefont {Zi{\k{e}}ba-Ost{\'o}j}, \citenamefont
  {Wasiluk}, \citenamefont {Ga{\'a}l}, \citenamefont {Sakanas}, \citenamefont
  {Burakowski}, \citenamefont {Mrowi{\'n}ski}, \citenamefont {Krajnik},
  \citenamefont {Xiong} \emph {et~al.}}]{holewa2024high}%
  \BibitemOpen
  \bibfield  {author} {\bibinfo {author} {\bibfnamefont {P.}~\bibnamefont
  {Holewa}}, \bibinfo {author} {\bibfnamefont {D.~A.}\ \bibnamefont {Vajner}},
  \bibinfo {author} {\bibfnamefont {E.}~\bibnamefont {Zi{\k{e}}ba-Ost{\'o}j}},
  \bibinfo {author} {\bibfnamefont {M.}~\bibnamefont {Wasiluk}}, \bibinfo
  {author} {\bibfnamefont {B.}~\bibnamefont {Ga{\'a}l}}, \bibinfo {author}
  {\bibfnamefont {A.}~\bibnamefont {Sakanas}}, \bibinfo {author} {\bibfnamefont
  {M.}~\bibnamefont {Burakowski}}, \bibinfo {author} {\bibfnamefont
  {P.}~\bibnamefont {Mrowi{\'n}ski}}, \bibinfo {author} {\bibfnamefont
  {B.}~\bibnamefont {Krajnik}}, \bibinfo {author} {\bibfnamefont
  {M.}~\bibnamefont {Xiong}}, \emph {et~al.},\ }\bibfield  {title} {\bibinfo
  {title} {High-throughput quantum photonic devices emitting indistinguishable
  photons in the telecom c-band},\ }\href
  {https://doi.org/10.1038/s41467-024-47551-7} {\bibfield  {journal} {\bibinfo
  {journal} {Nature Communications}\ }\textbf {\bibinfo {volume} {15}},\
  \bibinfo {pages} {3358} (\bibinfo {year} {2024})}\BibitemShut {NoStop}%
\bibitem [{\citenamefont {Yeung}\ \emph
  {et~al.}(2023{\natexlab{b}})\citenamefont {Yeung}, \citenamefont {Northeast},
  \citenamefont {Jin}, \citenamefont {Laferri{\`e}re}, \citenamefont
  {Korkusinski}, \citenamefont {Poole}, \citenamefont {Williams},\ and\
  \citenamefont {Dalacu}}]{Yeung_2023}%
  \BibitemOpen
  \bibfield  {author} {\bibinfo {author} {\bibfnamefont {E.}~\bibnamefont
  {Yeung}}, \bibinfo {author} {\bibfnamefont {D.~B.}\ \bibnamefont
  {Northeast}}, \bibinfo {author} {\bibfnamefont {J.}~\bibnamefont {Jin}},
  \bibinfo {author} {\bibfnamefont {P.}~\bibnamefont {Laferri{\`e}re}},
  \bibinfo {author} {\bibfnamefont {M.}~\bibnamefont {Korkusinski}}, \bibinfo
  {author} {\bibfnamefont {P.~J.}\ \bibnamefont {Poole}}, \bibinfo {author}
  {\bibfnamefont {R.~L.}\ \bibnamefont {Williams}},\ and\ \bibinfo {author}
  {\bibfnamefont {D.}~\bibnamefont {Dalacu}},\ }\bibfield  {title} {\bibinfo
  {title} {On-chip indistinguishable photons using iii-v nanowire/sin hybrid
  integration},\ }\href {https://doi.org/10.1103/physrevb.108.195417}
  {\bibfield  {journal} {\bibinfo  {journal} {Physical Review B}\ }\textbf
  {\bibinfo {volume} {108}},\ \bibinfo {pages} {195417} (\bibinfo {year}
  {2023}{\natexlab{b}})}\BibitemShut {NoStop}%
\bibitem [{\citenamefont {Bornadel}\ \emph {et~al.}(2024)\citenamefont
  {Bornadel}, \citenamefont {Alavijeh}, \citenamefont {Rasekh}, \citenamefont
  {Kamel}, \citenamefont {Asadi}, \citenamefont {Saglamyurek}, \citenamefont
  {Oblak},\ and\ \citenamefont {Simon}}]{bornadel2024hole}%
  \BibitemOpen
  \bibfield  {author} {\bibinfo {author} {\bibfnamefont {M.}~\bibnamefont
  {Bornadel}}, \bibinfo {author} {\bibfnamefont {S.~S.}\ \bibnamefont
  {Alavijeh}}, \bibinfo {author} {\bibfnamefont {F.}~\bibnamefont {Rasekh}},
  \bibinfo {author} {\bibfnamefont {N.~G.}\ \bibnamefont {Kamel}}, \bibinfo
  {author} {\bibfnamefont {F.~K.}\ \bibnamefont {Asadi}}, \bibinfo {author}
  {\bibfnamefont {E.}~\bibnamefont {Saglamyurek}}, \bibinfo {author}
  {\bibfnamefont {D.}~\bibnamefont {Oblak}},\ and\ \bibinfo {author}
  {\bibfnamefont {C.}~\bibnamefont {Simon}},\ }\bibfield  {title} {\bibinfo
  {title} {Hole burning experiments and modeling in erbium-doped silica glass
  fibers down to millikelvin temperatures: evidence for ultra-long population
  storage},\ }\href {https://doi.org/10.48550/arXiv.2412.16013} {\bibfield
  {journal} {\bibinfo  {journal} {arXiv preprint}\ } (\bibinfo {year}
  {2024})}\BibitemShut {NoStop}%
\bibitem [{\citenamefont {Saglamyurek}\ \emph
  {et~al.}(2015{\natexlab{b}})\citenamefont {Saglamyurek}, \citenamefont
  {Lutz}, \citenamefont {Veissier}, \citenamefont {Hedges}, \citenamefont
  {Thiel}, \citenamefont {Cone},\ and\ \citenamefont
  {Tittel}}]{Saglamyurek_2015_hole}%
  \BibitemOpen
  \bibfield  {author} {\bibinfo {author} {\bibfnamefont {E.}~\bibnamefont
  {Saglamyurek}}, \bibinfo {author} {\bibfnamefont {T.}~\bibnamefont {Lutz}},
  \bibinfo {author} {\bibfnamefont {L.}~\bibnamefont {Veissier}}, \bibinfo
  {author} {\bibfnamefont {M.~P.}\ \bibnamefont {Hedges}}, \bibinfo {author}
  {\bibfnamefont {C.~W.}\ \bibnamefont {Thiel}}, \bibinfo {author}
  {\bibfnamefont {R.~L.}\ \bibnamefont {Cone}},\ and\ \bibinfo {author}
  {\bibfnamefont {W.}~\bibnamefont {Tittel}},\ }\bibfield  {title} {\bibinfo
  {title} {Efficient and long-lived zeeman-sublevel atomic population storage
  in an erbium-doped glass fiber},\ }\href
  {https://doi.org/10.1103/physrevb.92.241111} {\bibfield  {journal} {\bibinfo
  {journal} {Physical Review B}\ }\textbf {\bibinfo {volume} {92}},\ \bibinfo
  {pages} {241111} (\bibinfo {year} {2015}{\natexlab{b}})}\BibitemShut
  {NoStop}%
\bibitem [{\citenamefont {Hahn}(1950)}]{hahn1950spin}%
  \BibitemOpen
  \bibfield  {author} {\bibinfo {author} {\bibfnamefont {E.~L.}\ \bibnamefont
  {Hahn}},\ }\bibfield  {title} {\bibinfo {title} {Spin echoes},\ }\href
  {https://doi.org/10.1103/PhysRev.80.580} {\bibfield  {journal} {\bibinfo
  {journal} {Physical review}\ }\textbf {\bibinfo {volume} {80}},\ \bibinfo
  {pages} {580} (\bibinfo {year} {1950})}\BibitemShut {NoStop}%
\bibitem [{\citenamefont {Li}\ \emph {et~al.}(2025)\citenamefont {Li},
  \citenamefont {Lei}, \citenamefont {Kling},\ and\ \citenamefont
  {Hosseini}}]{li-2025}%
  \BibitemOpen
  \bibfield  {author} {\bibinfo {author} {\bibfnamefont {Z.}~\bibnamefont
  {Li}}, \bibinfo {author} {\bibfnamefont {Y.}~\bibnamefont {Lei}}, \bibinfo
  {author} {\bibfnamefont {T.}~\bibnamefont {Kling}},\ and\ \bibinfo {author}
  {\bibfnamefont {M.}~\bibnamefont {Hosseini}},\ }\bibfield  {title} {\bibinfo
  {title} {{Efficient storage of multidimensional telecom photons in a
  Solid-State quantum memory}},\ }\bibfield  {journal} {\bibinfo  {journal}
  {Optica Quantum}\ }\href {https://doi.org/10.1364/opticaq.564321}
  {10.1364/opticaq.564321} (\bibinfo {year} {2025})\BibitemShut {NoStop}%
\bibitem [{\citenamefont {Usmani}\ \emph {et~al.}(2010)\citenamefont {Usmani},
  \citenamefont {Afzelius}, \citenamefont {De~Riedmatten},\ and\ \citenamefont
  {Gisin}}]{usmani-2010}%
  \BibitemOpen
  \bibfield  {author} {\bibinfo {author} {\bibfnamefont {I.}~\bibnamefont
  {Usmani}}, \bibinfo {author} {\bibfnamefont {M.}~\bibnamefont {Afzelius}},
  \bibinfo {author} {\bibfnamefont {H.}~\bibnamefont {De~Riedmatten}},\ and\
  \bibinfo {author} {\bibfnamefont {N.}~\bibnamefont {Gisin}},\ }\bibfield
  {title} {\bibinfo {title} {{Mapping multiple photonic qubits into and out of
  one solid-state atomic ensemble}},\ }\bibfield  {journal} {\bibinfo
  {journal} {Nature Communications}\ }\textbf {\bibinfo {volume} {1}},\ \href
  {https://doi.org/10.1038/ncomms1010} {10.1038/ncomms1010} (\bibinfo {year}
  {2010})\BibitemShut {NoStop}%
\bibitem [{\citenamefont {G{\"u}ndo{\u{g}}an}\ \emph
  {et~al.}(2012)\citenamefont {G{\"u}ndo{\u{g}}an}, \citenamefont {Ledingham},
  \citenamefont {Almasi}, \citenamefont {Cristiani},\ and\ \citenamefont
  {De~Riedmatten}}]{gundougan2012quantum}%
  \BibitemOpen
  \bibfield  {author} {\bibinfo {author} {\bibfnamefont {M.}~\bibnamefont
  {G{\"u}ndo{\u{g}}an}}, \bibinfo {author} {\bibfnamefont {P.~M.}\ \bibnamefont
  {Ledingham}}, \bibinfo {author} {\bibfnamefont {A.}~\bibnamefont {Almasi}},
  \bibinfo {author} {\bibfnamefont {M.}~\bibnamefont {Cristiani}},\ and\
  \bibinfo {author} {\bibfnamefont {H.}~\bibnamefont {De~Riedmatten}},\
  }\bibfield  {title} {\bibinfo {title} {Quantum storage of a photonic
  polarization qubit in a solid},\ }\href
  {https://doi.org/10.1103/PhysRevLett.108.190504} {\bibfield  {journal}
  {\bibinfo  {journal} {Physical review letters}\ }\textbf {\bibinfo {volume}
  {108}},\ \bibinfo {pages} {190504} (\bibinfo {year} {2012})}\BibitemShut
  {NoStop}%
\bibitem [{\citenamefont {Laferriere}\ \emph {et~al.}(2021)\citenamefont
  {Laferriere}, \citenamefont {Yeung}, \citenamefont {Korkusinski},
  \citenamefont {Poole}, \citenamefont {Williams}, \citenamefont {Dalacu},
  \citenamefont {Manalo}, \citenamefont {Cygorek}, \citenamefont {Altintas},\
  and\ \citenamefont {Hawrylak}}]{Laferriere2021}%
  \BibitemOpen
  \bibfield  {author} {\bibinfo {author} {\bibfnamefont {P.}~\bibnamefont
  {Laferriere}}, \bibinfo {author} {\bibfnamefont {E.}~\bibnamefont {Yeung}},
  \bibinfo {author} {\bibfnamefont {M.}~\bibnamefont {Korkusinski}}, \bibinfo
  {author} {\bibfnamefont {P.~J.}\ \bibnamefont {Poole}}, \bibinfo {author}
  {\bibfnamefont {R.~L.}\ \bibnamefont {Williams}}, \bibinfo {author}
  {\bibfnamefont {D.}~\bibnamefont {Dalacu}}, \bibinfo {author} {\bibfnamefont
  {J.}~\bibnamefont {Manalo}}, \bibinfo {author} {\bibfnamefont
  {M.}~\bibnamefont {Cygorek}}, \bibinfo {author} {\bibfnamefont
  {A.}~\bibnamefont {Altintas}},\ and\ \bibinfo {author} {\bibfnamefont
  {P.}~\bibnamefont {Hawrylak}},\ }\bibfield  {title} {\bibinfo {title}
  {Systematic study of the emission spectra of nanowire quantum dots},\
  }\bibfield  {journal} {\bibinfo  {journal} {Applied Physics Letters}\
  }\textbf {\bibinfo {volume} {118}},\ \href
  {https://doi.org/10.1063/5.0045880} {10.1063/5.0045880} (\bibinfo {year}
  {2021})\BibitemShut {NoStop}%
\bibitem [{\citenamefont {Laferriere}\ \emph {et~al.}(2020)\citenamefont
  {Laferriere}, \citenamefont {Yeung}, \citenamefont {Giner}, \citenamefont
  {Haffouz}, \citenamefont {Lapointe}, \citenamefont {Aers}, \citenamefont
  {Poole}, \citenamefont {Williams},\ and\ \citenamefont
  {Dalacu}}]{laferriere2020multiplexed}%
  \BibitemOpen
  \bibfield  {author} {\bibinfo {author} {\bibfnamefont {P.}~\bibnamefont
  {Laferriere}}, \bibinfo {author} {\bibfnamefont {E.}~\bibnamefont {Yeung}},
  \bibinfo {author} {\bibfnamefont {L.}~\bibnamefont {Giner}}, \bibinfo
  {author} {\bibfnamefont {S.}~\bibnamefont {Haffouz}}, \bibinfo {author}
  {\bibfnamefont {J.}~\bibnamefont {Lapointe}}, \bibinfo {author}
  {\bibfnamefont {G.~C.}\ \bibnamefont {Aers}}, \bibinfo {author}
  {\bibfnamefont {P.~J.}\ \bibnamefont {Poole}}, \bibinfo {author}
  {\bibfnamefont {R.~L.}\ \bibnamefont {Williams}},\ and\ \bibinfo {author}
  {\bibfnamefont {D.}~\bibnamefont {Dalacu}},\ }\bibfield  {title} {\bibinfo
  {title} {Multiplexed single-photon source based on multiple quantum dots
  embedded within a single nanowire},\ }\href
  {https://doi.org/10.1021/acs.nanolett.0c00607} {\bibfield  {journal}
  {\bibinfo  {journal} {Nano letters}\ }\textbf {\bibinfo {volume} {20}},\
  \bibinfo {pages} {3688} (\bibinfo {year} {2020})}\BibitemShut {NoStop}%
\end{thebibliography}%
\FloatBarrier

\end{document}